%% file: 2017-decay-magnetic-dipole.tex
\documentclass[aps,prb,twocolumn,showpacs,floatfix,a4paper]{revtex4}

\usepackage[utf8]{inputenc}
\usepackage[T1]{fontenc}

\usepackage{graphicx}
\usepackage{tabularx}
\usepackage{color}

\usepackage{amsmath}
\usepackage{amsfonts}
\usepackage{amssymb}
\usepackage{amstext, mathrsfs, textcomp}

\usepackage[colorlinks=true, citecolor=blue, urlcolor=blue, linkcolor=blue]{hyperref}

\usepackage[final]{changes}

\graphicspath{{Figures/}}

\newcommand{\TITLE}{Decay Rate of Magnetic Dipoles near Non--magnetic Nanostructures}

\newcommand{\ADDRESSCEMES}{CEMES-CNRS, Universit\'e de Toulouse, CNRS, UPS, Toulouse, France}

\hypersetup{
	pdftitle = {\TITLE},
	pdfsubject = {},
	pdfauthor = {Peter R. Wiecha et al.},
	pdfkeywords = {},
	pdfcreator = {LaTeX},
	pdfproducer = {LaTeX with hyperref and thumbpdf}
}

\begin{document}

%

\title{\TITLE}

\author{\firstname{Peter R.} \surname{Wiecha}}
\email[e-mail~: ]{peter.wiecha@cemes.fr}
\affiliation{\ADDRESSCEMES}

\author{\firstname{Arnaud} \surname{Arbouet}}
\affiliation{\ADDRESSCEMES}

\author{\firstname{Aur\'{e}lien} \surname{Cuche}}
\affiliation{\ADDRESSCEMES}

\author{\firstname{Vincent} \surname{Paillard}}
\affiliation{\ADDRESSCEMES}

\author{\firstname{Christian} \surname{Girard}}
\affiliation{\ADDRESSCEMES}

\begin{abstract}
In this article, we propose a concise theoretical framework based on mixed field-susceptibilities to describe the decay of magnetic dipoles induced by non--magnetic nanostructures.
This approach is first illustrated in simple cases in which analytical expressions of the decay rate can be obtained.
We then show that a more refined numerical implementation of this formalism involving a volume discretization and the computation of a generalized propagator can predict the dynamics of magnetic dipoles in the vicinity of nanostructures of arbitrary geometries.
We finally demonstrate the versatility of this numerical method by coupling it to an evolutionary optimization algorithm. 
In this way we predict a structure geometry which maximally promotes the decay of magnetic transitions with respect to electric emitters.
\end{abstract}
\maketitle
\section{Introduction}

During the last two decades, the development of nano-optics has provided a wealth of strategies to tailor electric and magnetic fields down to the subwavelength scale
\cite{novotny_principles_2006}.
In particular, optical nano-antennas have allowed to modify the intensity, dynamics or directionality of light emission from fluorophores placed in the near-field of nano-objects using concepts from the radiofrequency domain
\cite{anger_enhancement_2006, kuhn_enhancement_2006, kinkhabwala_large_2009, curto_unidirectional_2010, biagioni_nanoantennas_2012, novotny_antennas_2011, klimov_radiative_1996}.
These studies have been performed nearly exclusively on fluorophores supporting electric dipole (ED) transitions, the latter being $\sim a_{0}/\lambda_0 \sim 10^{4}-10^5$ larger than their magnetic dipole (MD) counterpart in the optical frequency range ($a_{0}$ being the Bohr radius and $\lambda_0$ the transition wavelength)
\cite{giessen_glimpsing_2009}.
Recently, delicate experiments have addressed light emission from
rare earth doped emitters supporting both strong MD and ED transitions
\cite{aigouy_mapping_2014,choi_selective_2016}. 
Three--dimensional maps of the luminescence of Eu$^{3+}$ doped nanocrystals scanned in the near-field of gold stripes
have revealed variations in the relative intensities of ED and MD transitions
\cite{karaveli_spectral_2011, aigouy_mapping_2014}.
In these experiments, the fluorescence intensity, photon statistics and branching ratios are directly related to the decay rates of the ED or MD radiative transitions, 
the latter being ultimately connected to the electric or magnetic part of the local density of electromagnetic states (EM-LDOS)
\cite{aigouy_mapping_2014, carminati_electromagnetic_2015, baranov_modifying_2017, cuche_near-field_2017}.
Independently of the nature of the transition, the alteration of the EM-LDOS by a nano\-structure arises from the 
back action of the electric or magnetic field on the transition dipole \cite{purcell_resonance_1946, chicanne_imaging_2002, anger_enhancement_2006, rolly_promoting_2012, chigrin_emission_2016}.

Analytical expressions for the decay of magnetic transitions have been derived for the simple case of single \cite{chew_fluorescent_1979, klimov_electric_2005, schmidt_dielectric_2012} or also multiple spheres \cite{stout_multipole_2011, rolly_promoting_2012}.
For more complex geometries or arrangements of nano-structures, standard numerical tools like finite difference time domain (FDTD) or finite elements method (FEM) can be employed to calculate magnetic decay rates.
To do so, a kind of numerical experiment is performed where the radiated power of a dipole emitter is compared for the cases with, and in absence of a nanostructure. \cite{feng_controlling_2011, hein_tailoring_2013, albella_low-loss_2013, mivelle_strong_2015, feng_all-dielectric_2016}

Whereas, the underlying physics is well understood, a unified description of the dynamics of a fluorophore supporting MD transitions in the presence of non-magnetic nanostructures of arbitrary shape is still lacking.

The confinement of the magnetic field around non-magnetic nano-objects
arises from the spatial variations of the electric near--field in the immediate proximity of
a nanostructure.
When the surface is illuminated by a plane wave or an evanescent surface wave,
both experimental data and numerical simulations
reveal spatial modulations in the electric and magnetic near-field intensities.
For example, the magnetic field intensity recorded above subwavelength sized dielectric particles,  
excited by a p--polarized surface wave, has a strong and dark contrast
while a completely opposite behavior is
observed for the electric field intensity
\cite{burresi_probing_2009,devaux_detection_2000,devaux_local_2000}.
If now, the nanostructure is illuminated, no longer
by a plane wave but by a dipole source, the response fields (electric or magnetic) are different and shape the decay rate and the corresponding dipolar luminescence. 
From a mathematical point of view, the magnetic near-field can be described by
a set of mixed field--susceptibilities capable of connecting an electrical polarization, oscillating
at an optical frequency $\omega_{0}$, to a magnetic field vector 
oscillating at the same frequency
\cite{girard_optical_1997, schroter_modelling_2003}.
In fact, these field--susceptibilities are a generalized form 
of the usual Green dyadic tensor
\cite{martin_generalized_1995, girard_near_2005}. 
Historically, they were introduced by G.S. Agarwal to describe energy transfers
in the presence of dielectric or metallic planar surfaces
\cite{agarwal_quantum_1975}.
Mixed field--susceptibilities can be used to evaluate the optical magnetic near-field or the optical response of nano-structures possessing an intrinsic magnetic polarizability, like metallic rings or split-rings. \cite{schroter_modelling_2003, sersic_magnetoelectric_2011}
In recent works, they have been used to separately study the magnetic and electric part of the LDOS close to a surface
\cite{kwadrin_probing_2013} and for the calculation of the EM-LDOS in proximity of periodic arrays of magneto-electric point scatterers \cite{lunnemann_local_2016}.

\begin{figure}
\centering\includegraphics[angle =0.]{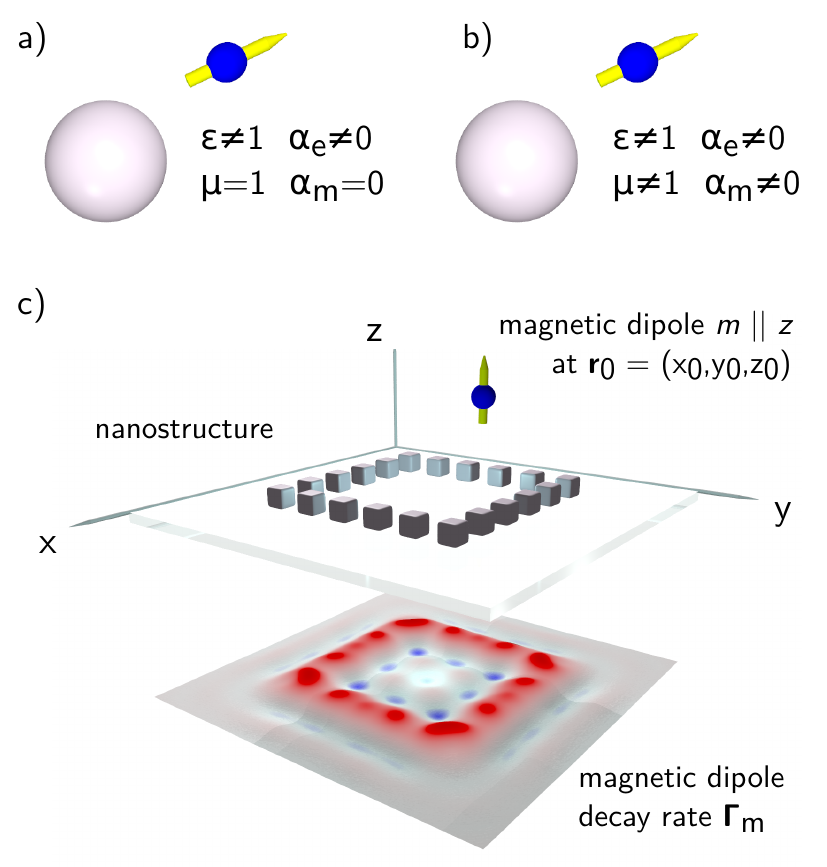}
\caption{(color online) 
a) and b) Illustration of two possible kinds of optical coupling between an oscillating magnetic dipole and a nanostructure defined by arbitrary optical permittivity $\epsilon(\omega_{0})$ and permeability $\mu(\omega_{0})$, with \(\alpha_e\) and \(\alpha_m\) the electric and magnetic polarizabilities.
The present study focuses on the first case in which the magnetic dipole is coupled to a non-magnetic nanostructure, as shown in a).
c) Decay rate of a magnetic dipole $\textit{m}(\omega_{0})$ oriented along $0Z$ and scanned at $z_{0}=30\,$nm above the depicted nanostructure.
}
\label{TAB-MAG}
\end{figure}

In this article, we first extend Agarwal's theory by presenting a new analytical scheme yielding the total decay rate of a MD transition $\Gamma_{m}$ in terms of 
mixed {\it electric--magnetic} field--susceptibilities.
From this concise mathematical framework, we develop a flexible  and powerful numerical tool 
to compute the decay rate of magnetic dipoles near dielectric or metallic nanostructures of arbitrary shapes (see example in Fig. \ref{TAB-MAG}).
In a second step, we explore the decay rate maps generated by the coupling between rare--earth atoms and dielectric nanostructures. 
We highlight and discuss the differences between electric and magnetic decay rate topographies.
Finally, we demonstrate the versatility of our mathematical framework by coupling it to an evolutionary optimization algorithm to predict a metallic nano-structure yielding an optimum contrast between the magnetic and electric parts of the EM-LDOS.

\section{Magnetic field-susceptibility for non-magnetic structures}

The two possible kinds of coupling between a magnetic dipole transition
and a subwavelength sized sphere are schematized in figure~\ref{TAB-MAG}-a and b. 
The first one is the coupling with a standard material (bulk metal, dielectric or semiconductor)
which does not possess any intrinsic magnetic response (i.e. for which the magnetic permeability is equal to unity (CGS units))
while the second one is the direct magnetic coupling such as the one involved in the
presence of artificial {\it left--handed} materials, i.e materials with simultaneously negative permeability and permittivity \cite{pendry_negative_2000, soukoulis_negative_2007}.
We address exclusively the first situation and therefore assume that $\mu(\omega_{0})$ = 1
at all wavelengths. 
We consider the geometry depicted in figure~\ref{TAB-MAG}.
The electric and magnetic fields generated at ${\bf r}$ by a magnetic dipole ${\bf m}(\omega_{0})$ located at 
${\bf r}_{0}$ are defined by \cite{agarwal_quantum_1975}: 
\begin{equation}
{\bf E}_{0}({\bf r},\omega_{0})=ik_{0}\nabla_{{\bf r}}\wedge
{\cal G}_{0}({\bf r},{\bf r}_{0},\omega_{0})\cdot{\bf m}(\omega_{0})
\label{E0M}
\; ,
\end{equation}
and
\begin{equation}
{\bf H}_{0}({\bf r},\omega_{0})=\Big\{{\bf I}k_{0}^{2}+\nabla_{{\bf r}}\nabla_{{\bf r}}
\Big\}
{\cal G}_{0}({\bf r},{\bf r}_{0},\omega_{0})\cdot{\bf m}(\omega_{0})
\; ,
\label{H0M}
\end{equation}
where $k_{0}=2\pi / \lambda_0$ is the wave vector in vacuum and ${\cal G}_{0}({\bf r},{\bf r}_{0},\omega_{0})$
= $\exp (ik_{0}|{\bf r}-{\bf r}_{0}|)/|{\bf r}-{\bf r}_{0}|$
represents the scalar Green function.
From these two equations, we can define two field susceptibilities:
\begin{equation}
{\bf E}_{0}({\bf r},\omega_{0})={\cal S}^{EH}({\bf r},{\bf r}_{0},\omega_{0})\cdot{\bf m}(\omega_{0}) 
\; ,
\end{equation}
and
\begin{equation}{\bf H}_{0}({\bf r},\omega_{0})={\cal S}^{HH}({\bf r},{\bf r}_{0},\omega_{0})\cdot{\bf m}(\omega_{0})
\end{equation}
in which the dyadic tensors ${\cal S}^{EH}({\bf r},{\bf r}_{0},\omega_{0})$ and ${\cal S}^{HH}({\bf r},{\bf r}_{0},\omega_{0})$
are constructed by identification with equations (\ref{E0M}) and (\ref{H0M}).
For the mixed dyad ${\cal S}^{EH}({\bf r},{\bf r}_{0},\omega_{0})$ this identification yields the expression of the nine analytical components:
\begin{eqnarray}
{\cal S}^{EH}({\bf r},{\bf r}_{0},\omega_{0})=ik_{0}
\left (
\begin{array}{ccc}
0 & -\frac{\partial {\cal G}_{0}}{\partial z} & \frac{\partial {\cal G}_{0}}{\partial y} \\
\\
\frac{\partial {\cal G}_{0}}{\partial z} & 0 & -\frac{\partial {\cal G}_{0}}{\partial x} \\
\\
-\frac{\partial {\cal G}_{0}}{\partial y} & \frac{\partial {\cal G}_{0}}{\partial x} & 0
\end{array}
\right)
\; .
\end{eqnarray}
Equations (\ref{E0M}) and (\ref{H0M}) define the {\it so--called} illumination
field. 
Since the materials considered in this article 
do not directly respond to the optical magnetic field, 
the coupling with the nanoparticle is entirely described by the first equation. 
A complete theoretical investigation of this illumination mode requires
the accurate computation of the optical field distribution inside the nanostructure 
for every location ${\bf r}_{0}$ of the magnetic dipole.
As discussed in the literature, the recent developments of real space 
approaches for electromagnetic scattering and light confinement 
established powerful tools for the calculation of the
electromagnetic response of complex mesoscopic systems to arbitrary
illumination field \cite{girard_near_2005}.
Particularly, the technique of the generalized field propagator described in 
reference \cite{martin_generalized_1995}, provides a convenient basis to derive the 
electromagnetic response of an  arbitrary system to a great number of different external excitation fields
\cite{arbouet_electron_2014}.
Our approach is based on the computation
of a unique generalized field propagator ${\cal K}({\bf r}',{\bf r''},\omega_0)$
that contains the entire response of the nanostructure to any 
incident electric
field ${\bf E}_{0}({\bf r}'',\omega_{0})$.
Consequently, the self-consistent electric field ${\bf E}({\bf r}_{0},{\bf r}',\omega_{0})$ created inside the nanosystem by a magnetic dipole located at ${\bf r}_{0}$, 
can be written as: 
\begin{equation}
\label{KD}
{\bf E}({\bf r}_{0},{\bf r}',\omega_{0})=\int_{v}{\cal K}({\bf r}',{\bf r}'',\omega_0)\cdot
{\bf E_{0}}({\bf r}'',\omega_0) {\text d}{\bf r}'' \,,
\end{equation} 
in which the integral runs over 
the volume $v$ of the particle.
As demonstrated in reference~\onlinecite{martin_generalized_1995}, the dyadic ${\cal K}$ writes
\begin{equation}
{\cal K}({\bf r}',{\bf r}'',\omega_{0}) = \delta({\bf r}'-{\bf r}'')+
{\cal S}({\bf r}',{\bf r}'',\omega_{0})\cdot\chi({\bf r}'',\omega_{0})\, ,
\end{equation}
where $\delta$ is the three-dimensional Dirac function, ${\cal S}({\bf r}',{\bf r}'',\omega_{0})$ is the optical field--susceptibility tensor of the nanostructure of electric susceptibility $\chi({\bf r}'',\omega_{0})$.

Equation (\ref{KD}) gives access to the electric field 
inside the nanostructure and therefore to the polarization
${\mathbf{P}}({\bf r}_{0},{\bf r}'',\omega_{0})= \chi({\bf r}'',\omega_{0})\cdot{\bf E}({\bf r}_{0},{\bf r}'',\omega_{0})$
induced for each position ${\bf r}_{0}$ of the magnetic dipole.
The magnetic field generated outside of the particle can then be calculated 
by introducing the second mixed field--susceptibility ${\cal S}^{HE}({\bf r},{\bf r}',\omega_{0})$
$=$ ${\cal S}^{EH}({\bf r}',{\bf r},\omega_{0})$ \cite{agarwal_quantum_1975}:
\begin{equation}
\label{HEXT}
{\bf H}({\bf r}_{0},{\bf r},\omega_{0})
=\int_{v}{\cal S}^{HE}({\bf r},{\bf r}',\omega_{0})\cdot{\mathbf{P}}({\bf r}_{0},{\bf r}',\omega_{0})
d{\bf r}' \; ,
\end{equation}
which, in a concise form, leads to:
\begin{equation}
{\bf H}({\bf r}_{0},{\bf r},\omega_{0})={\cal S}_{p}^{HH}({\bf r},{\bf r}_{0},\omega_{0})\cdot
{\bf m}(\omega_{0})
\; ,
\end{equation}
where ${\cal S}_{p}^{HH}({\bf r},{\bf r}_{0},\omega_{0})$ defines the magnetic field susceptibility associated
with the nanostructure $(p)$:
\begin{eqnarray}
{\cal S}_{p}^{HH}({\bf r},{\bf r}_{0},\omega_{0})=
\int_{v}d{\bf r}'\int_{v}d{\bf r}''
{\cal S}^{HE}({\bf r},{\bf r}',\omega_{0})
\nonumber
\\
\cdot \chi({\bf r}',\omega_{0}) \cdot {\cal K}({\bf r}',{\bf r}'',\omega_{0})\cdot{\cal S}^{EH}({\bf r}'',{\bf r}_{0},\omega_{0})
\; .
\label{SHHP}
\end{eqnarray}
Here, the dot ``\(\cdot\)'' signifies the matrix product.
This general relationship, derived from the theory of linear response, brings to light the complex link between the electrical response of matter (contained in~\(\chi\) and~\({\cal K}\)) and the magnetic response of vacuum, through the mixed propagators \({\cal S}^{EH}\) and \({\cal S}^{HE}\).
The combination of these response functions shows in a concise way
how a nano-structure, which originally does not possess any magnetic response in the optical spectrum,
can nevertheless yield a {\it magnetic--magnetic} response.
Equation~\eqref{SHHP} summarizes with mathematical clarity the back-action of the electromagnetic near-field on a magnetic quantum emitter via the curl of the electric field, mediated by the presence of a non-magnetic nanostructure.


\section{Magnetic dipole decay-rate close to small dielectric particles}

Equation~\eqref{SHHP} allows us to obtain a general expression for the decay rate $\Gamma_{m}({\bf r}_{0},\omega_{0})$
associated with a {\it magnetic dipole transition} of amplitude $m_{eg}$ 
\cite{carminati_electromagnetic_2015}:
%
\begin{eqnarray}
\Gamma_{m}({\bf r}_{0},\omega_{0})=\Gamma_{m}^{0}(\omega_{0})
\nonumber
\\ 
\times \left\{1+
\frac{3}{2k_{0}^{3}}{\bf u}\cdot \text{Im}\big(
{\cal S}_{p}^{HH}({\bf r}_{0},{\bf r}_{0},\omega_{0})\big)\cdot{\bf u}
\right\} \; ,
\label{GGAMM}
\end{eqnarray} 
where $\Gamma_{m}^{0}(\omega_{0})$ = $ 4k_{0}^{3}m_{eg}^{2}/3\hbar$ represents the natural decay rate of
the magnetic transition and ${\bf u}$ labels the dipole orientation. 

\begin{figure}
\centering\includegraphics[angle =0.]{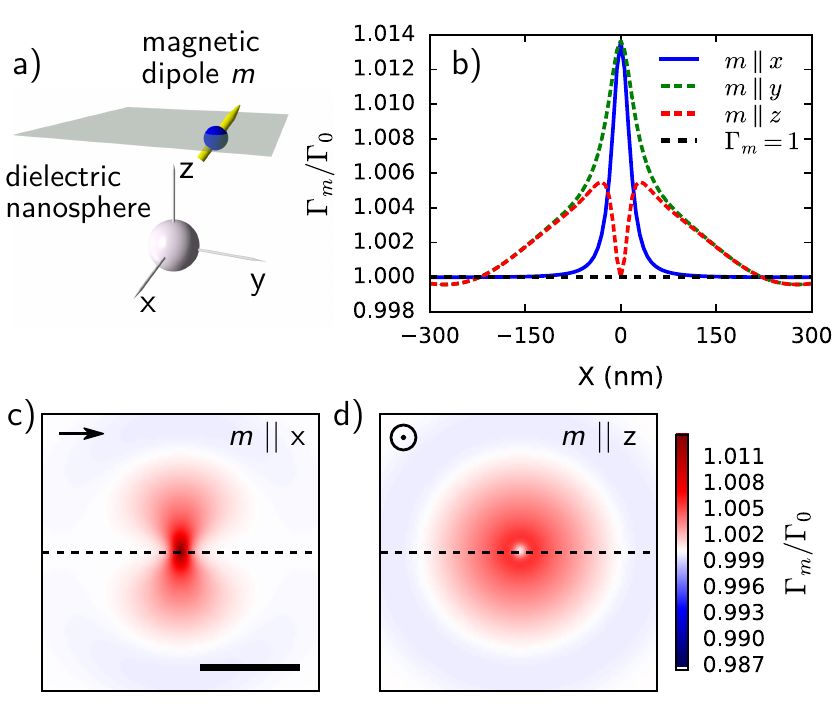}
\caption{(color online) (a) Single dielectric sphere of polarizability $\alpha_{e} = 1688$\,nm$^{3}$ (corresponding to $n=2$, $r=15\,$nm), raster scanned by a magnetic dipole at constant height $z_{0} = 20\,$nm.
(b) Cross-section of $\overline{\Gamma}_{m}$ computed at $\lambda_0=500\,$nm for  the three orientations $u_{x}$, $u_{y}$ and $u_{z}$ of the dipole. 
(c), (d) corresponding maps computed in the plane ($x_{0}$, $y_{0}$, $z_{0} = 20\,$nm). The maps have been computed from the complete expression of Equation~\eqref{SHHA}. 
Maps are \(600\times 600\,\)nm\(^2\), scalebar is \(200\,\)nm.}
\label{SINGLE}
\end{figure}

The next objective of this article is 
to supply a full analytical treatment of $\Gamma_{m}({\bf r}_{0},\omega_{0})$.
To achieve this goal, we deliberately reduce the physical model to a simple two--level system
coupled to a single spherical nanoparticle as shown in figure~\ref{SINGLE}-a. 
We have chosen to illustrate our method with dielectric materials as they offer an interesting alternative to metals with reduced dissipative losses and large resonant enhancement of both electric and magnetic near-fields
\cite{bakker_magnetic_2015, decker_resonant_2016, kuznetsov_optically_2016}.

In this case, a set of simple analytical equations can be derived that include
all the physical effects mentioned above. 
Indeed, we have ${\cal K}({\bf r}',{\bf r}'',\omega_{0})$ =
${\cal I}\delta({\bf r}'-{\bf r}'')$ (${\cal I}$ being the identity tensor),   
$\chi({\bf r}'',\omega_{0})$ = $\alpha_{e}(\omega_{0})\delta({\bf r}'')$, where $\alpha_{e}(\omega_{0})$
is the dynamical dipolar polarizability of the sphere, and finally:
\begin{equation}
{\cal S}_{p}^{HH}({\bf r},{\bf r}_{0},\omega_{0})=
\alpha_{e}(\omega_{0}){\cal S}^{HE}({\bf r},0,\omega_{0})\cdot{\cal S}^{EH}(0,{\bf r}_{0},\omega_{0})
\; .
\end{equation}
This relation can be further simplified
by replacing both ${\cal S}^{EH}$ and ${\cal S}^{HE}$ by their analytical expressions. 
In a plane defined by $x_{0}=0$, i.e. ${\bf r}_{0}$ = $(0,y_{0},z_{0})$, we get
the following simple expression when ${\bf r}$ = ${\bf r}_{0}$ (c.f. equation ~\eqref{GGAMM}):
\begin{equation}
\begin{split}
{\cal S}_{p}^{HH}({\bf r}_{0},{\bf r}_{0},\omega_{0}) = \\
\alpha_{e}(\omega_{0}){\cal A}e^{2ik_{0}r_{0}}
\left\{
 -\frac{k_{0}^{4}}{r_{0}^{4}}
 -\frac{2ik_{0}^{3}}{r_{0}^{5}}
 +\frac{k_{0}^{2}}{r_{0}^{6}}
\right\}
\; ,
\label{SHHA}
\end{split}
\end{equation}
where $r_0 = |\mathbf{r}_0|$ and the matrix ${\cal A}$ is defined by:
\begin{eqnarray}
{\cal A}=
\left (
\begin{array}{ccc}
y_{0}^{2}+z_{0}^{2} & 0 & 0 \\
\\
0 & z_{0}^{2} & -y_{0}z_{0} \\
\\
0 & -y_{0}z_{0} & y_{0}^{2} 
\end{array}
\right)
\; .
\label{AA}
\end{eqnarray}
In consequence, \({\cal S}_{p}^{HH}\) has the dimension of an inverse volume.\footnote{\(\alpha_e\) has a dimension of \(r^3\), \({\cal A}\) of \(r^2\) and all terms in the curly brackets of Eq.~\eqref{SHHA} are homogeneous to \(r^{-8}\)}
A concise expression of the normalized magnetic decay rate $\overline{\Gamma}_{m}$ = $\Gamma_{m}/\Gamma_{m}^{0}$
can then be deduced
by replacing this relation into (\ref{GGAMM}):   
%
\begin{equation}
\begin{split}
\overline{\Gamma}_{m}({\bf r}_{0},\omega_{0})=1+\alpha_{e}(\omega_{0})
{\bf u}\cdot{\cal A}\cdot{\bf u}
\\
\left\{ 
  \sin(2k_{0}r_{0})
     \left(
        -\frac{3k_{0}}{2r_{0}^{4}}
        +\frac{3}{2k_{0}r_{0}^{6}}
     \right)
  -\cos(2k_{0}r_{0})
    \frac{3}{r_{0}^{5}} 
\right\}
\; ,
\label{GAM-ANALY}
\end{split}
\end{equation}
in which the polarizability dissipation term $\textrm{Im} \, \alpha_{e}(\omega_{0})$ has been neglected.
We set $x_0 = 0$ to obtain the most simple equations possible. 
Adding it as free parameter is straightforward, yet renders the equations~\eqref{AA} and~\eqref{GAM-ANALY} more complex.
The case of a single dipolar dielectric sphere presented in figure~\ref{SINGLE} shows that the contrast patterns are extremely sensitive to the dipole orientation.  
The contrast is generally positive on top of the particle except when the dipole 
is aligned perpendicularly to the scanning plane ($x_{0}$,$y_{0}$) in which case it vanishes, the sphere becoming invisible for the magnetic dipole.
Such a peculiar behavior explicitly appears in equations~\eqref{AA} and~\eqref{GAM-ANALY} for small interaction distances, in particular, when the magnetic dipole enters the very subwavelength range corresponding to $2k_{0}r_{0}$ $\ll$ 1.
%
%
\begin{figure}
\centering\includegraphics[angle=0.]{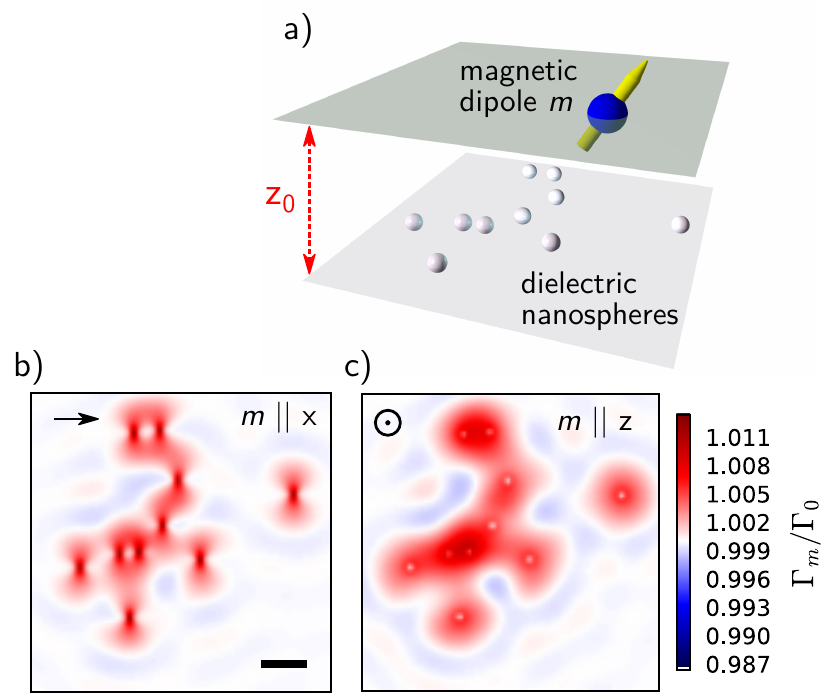}
\caption{(color online) (a) Random distribution of dielectric spheres
of polarizability $\alpha_{e} = 1688\,$nm$^{3}$ (corresponding to $n=2$, $r=15\,$nm), 
raster scanned by a magnetic dipole at constant height $z_{0} = 20\,$nm.
(b) and (c) show normalized decay maps for $u_{x}$ and $u_{z}$ orientation of the magnetic dipole, respectively.
Computed with $\lambda_0 = 500\,$nm.
Maps are \(1600\times 1600\,\)nm\(^2\), scalebar is \(200\,\)nm.}
\label{MANY}
\end{figure}
%
%
As a second example, we consider in figure~\ref{MANY} a set
of $p$ identical dielectric particles deposited on a transparent substrate
positioned at random locations ${\bf r}_{i}$ ($i$ = 1 to $p$).
The optical properties of such a system
can be described by first inserting the relation:
\begin{equation}
\chi({\bf r},\omega_{0}) = \alpha_{e}(\omega_{0})\sum_{i=1}^{p}\delta({\bf r}-{\bf r}_{i})
\; 
\label{CHI-MANY}
\end{equation}
in equation (\ref{SHHP}) and then in expression (\ref{GGAMM}).
The results are presented in figure~\ref{MANY}-b and~c.
When the particles are well-separated from each other, typically by one wavelength or more, 
they display a similar contrast as the one described in figure~\ref{SINGLE}.
This contrast is reinforced when several particles are grouped together.
Isolated particles and assemblies of particles
are surrounded by pseudoperiodic ripples that reveal the interferences 
between the emitting magnetic dipole and the sample.

\begin{figure}
\centering\includegraphics[width=0.95\linewidth,angle =0.]{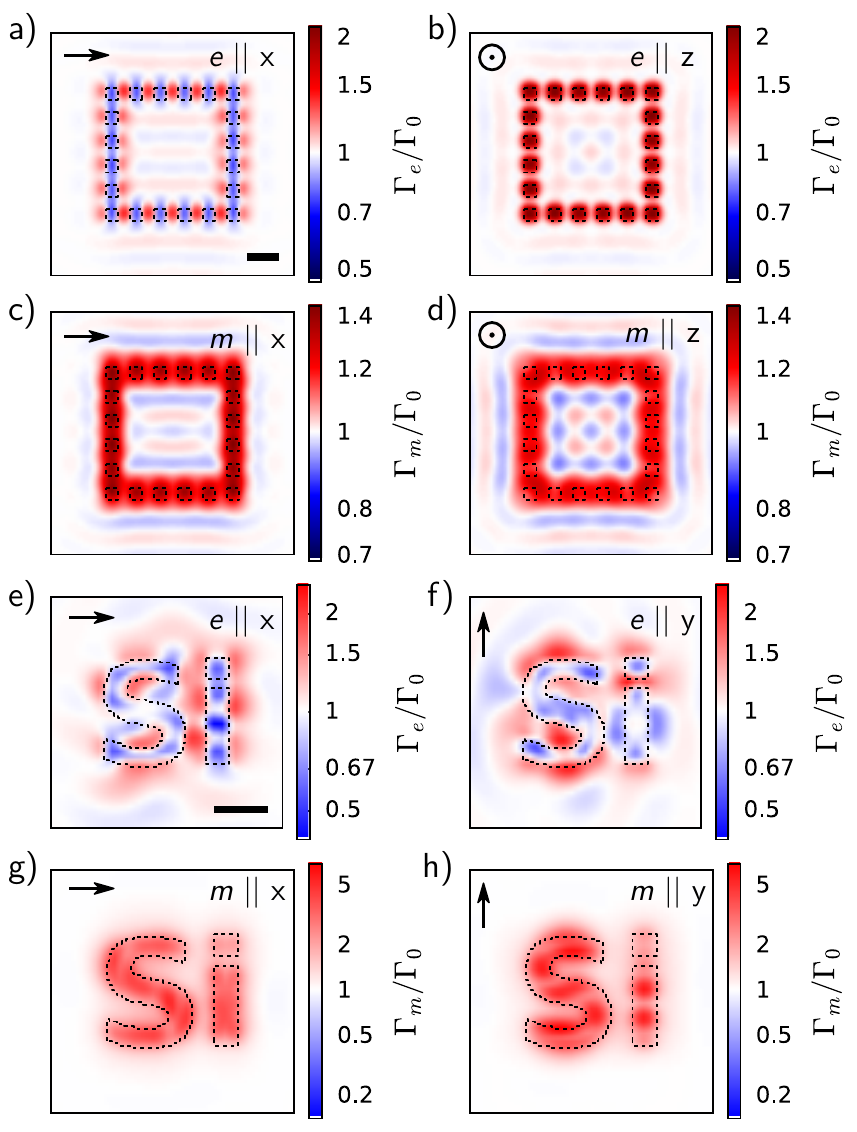}
\caption{(color online) 
(a-d) Maps of the decay rates of electric (ED) and magnetic dipoles (MD) \(30\,\)nm above a square corral of 20 dielectric nanocubes. 
The spacing between each cube of dimensions $100\times 100 \times 100\,\mathrm{nm}^{3}$ is $100\,$nm and the optical index is $n=2.0$. 
The maps (a-d) have a size of \(2\times 2\,\)\textmu m\(^2\).
(a) ED oriented along the $OX$ axis,
(b) ED oriented along the $OZ$ axis, (c) and (d) 
same computation for a MD.
(e-h) Maps of the ED (e-f) and MD (g-h) decay rate \(30\,\)nm above a silicon nanostructure (\(n = 4.3\)) composing the letters ``Si'' (structure height \(H=50\,\)nm).
The maps (e-h) are \(1\times 1\,\)\textmu m\(^2\) large.
(e) ED along $OX$, (f) ED along $OY$, (g) and (h) same for a MD.
A logarithmic colorscale is used due to the larger contrast in the decay rates.
All maps are computed at $\lambda_0 = 500\,$nm. 
Dashed black lines indicated the contours of the nanostructures.
Scale bars are \(200\,\)nm.}
\label{CORRAL}
\end{figure}

\section{Electric and magnetic dipole decay-rate close to complex dielectric nano-structures}

Whereas equations (\ref{SHHP}) and (\ref{GGAMM}) provide analytical expressions of the decay rate of magnetic dipoles placed close to very simple nano-objects, these equations can be complemented by an adequate 
discretization of the particle volume to describe light emission from dipoles in the vicinity of nanostructures of arbitrary geometries.
To this end, we numerically implement the complete computation of the generalized propapagtor ${\cal K}({\bf r}',{\bf r}'',\omega_0)$ as described in reference~\onlinecite{martin_generalized_1995}, together with $\chi({\bf r},\omega_{0})=(\epsilon(\omega_{0})-\epsilon_{\text{env}})/4\pi$ (CGS unit), where $\epsilon_{\text{env}}$ defines the permittivity of the environment.
We then use the propagator \({\cal K}\) associated to the nanostructure with the mixed-field susceptibilities \({\cal S}^{HE}\) and \({\cal S}^{EH}\) in a discretized version of equation~\eqref{SHHP}:
\begin{multline}
{\cal S}_{p}^{HH}({\bf r},{\bf r}_{0},\omega_{0})=
\sum\limits_{i=1}^N V_{\text{cell}}
\sum\limits_{j=1}^N V_{\text{cell}} \,
{\cal S}^{HE}({\bf r},{\bf r}_i,\omega_{0})
\\
\cdot \chi({\bf r}_i,\omega_{0}) \cdot {\cal K}({\bf r}_i,{\bf r}_j,\omega_{0})\cdot{\cal S}^{EH}({\bf r}_j,{\bf r}_{0},\omega_{0})
\; .
\label{SHHP_discretized}
\end{multline}
The sums (indexes \(i\) and \(j\)) run over all \(N\) discretization cells (of volume \(V_{\text{cell}}\)) forming the nanostructure.
This numerical procedure gives access to the optical response of complex systems, such as the ones described in figure~\ref{CORRAL}. 
In this example, we have applied this technique to visualize the footprint induced  
by a perfect square corral composed of 20 dielectric structures in the initially flat ED and MD decay rate maps. 
The extension of the entire nanostructure is \(1.1\,\)\textmu m, the refractive index is $n_{\text{pad}}=2$. 
A modification of the decay rates ranging between 20 and 50 $\%$ is obtained when the magnetic dipole is \(30\,\)nm above the nanostructures (fig.~\ref{CORRAL}c-d).
Although the coupling is more efficient with an electric dipole (fig.~\ref{CORRAL}a-b), especially when it is perpendicular, the coupling of the magnetic dipole
with the dielectric structure remains quite significant and could be easily observed. 
In particular, the normalized contrast $\overline{\Gamma}_{m}-1$ will be further enhanced when increasing $n_{\text{pad}}$ from 2 to 4 or 5 using high optical index dielectric or semiconductor materials (TiO$_2$, Si, or even Ge).
To demonstrate this enhancement, we show in fig.~\ref{CORRAL}e-h a flat silicon structure \(n \approx 4.3\) forming the letters ``Si'', on which a magnetic decay rate enhancement of more than a factor~\(5\) can be observed.
Moreover, we notice that the maps of the ED and MD decay rates display very specific features
that will allow to discriminate unambiguously the {\it electric} 
or {\it magnetic} nature of the atomic transition. 
A similar identification method has been proposed and demonstrated using back-focal plane imaging of electric / magnetic dipole luminescence from rare-earth-doped films.\cite{taminiau_quantifying_2012, li_quantifying_2014} 
Our results suggest an alternative discrimination technique using nano-structured substrates, which could be performed on less complex optical detection schemes.

\begin{figure*}
\centering\includegraphics[width=0.95\linewidth,angle =0.]{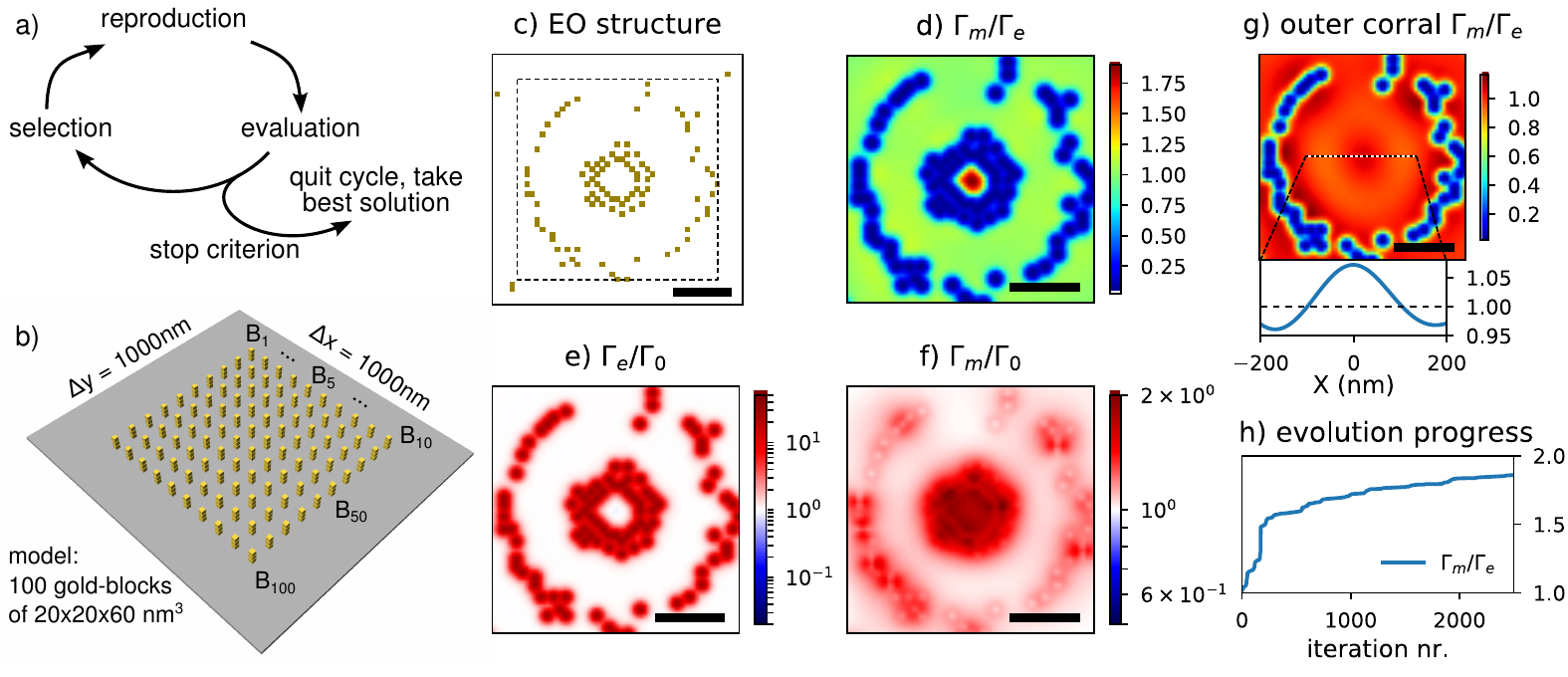}
\caption{(color online) 
(a) Evolutionary optimization cycle.
(b) Sketch of the structure model for optimization: Free parameters are the positions of \(100\) gold blocks (B\(_i\)) on the \(XY\) plane (in vacuum).
(c) Gold structure for optimum \(\Gamma_m / \Gamma_e\) contrast at the center (\(\mathbf{r}_0=(0,0,80)\)\,nm), found by EO. 
(d-g) Decay rate analysis of the EO solution. 
(d) Mapping of the ratio of magnetic and electric decay rate \(20\)\,nm above the structure.
(e-f) Relative electric and magnetic decay rates above the structure, respectively.
(g) Ratio of magnetic and electric decay rate for only the outer part of the structure, also leading to an enhancement of \(\Gamma_m\) at the target location. At the bottom of (g) \(\Gamma_m / \Gamma_e\) is shown along a profile in the center of the map.
(h) Progress of the evolutionary algorithm.
A logarithmic colorscale is used for the maps (e-f).
All results are computed at $\lambda_0 = 500\,$nm and for a dipole orientation along \(OZ\).
Scale bars are \(200\,\)nm. 
Mapping (c) is \(1000\times 1000\,\)nm\(^2\), (d-g) are \(800\times 800\,\)nm\(^2\) large (area indicated by a dashed square in c).}
\label{evolopt}
\end{figure*}

For instance, when the emitting dipole is oriented along the $(OX)$ axis (maps (a), (c), (e) and (g) of figure~\ref{CORRAL}), we observe a contrast reversal above the dielectric pads when passing from an {\it electric} to a {\it magnetic} dipole.
This striking phenomenon is accompanied by a shift of the fringe pattern inside the corral by half a wavelength.
Finally, another type of contrast change is observed when the dipole is perpendicular to the sample.
In this second case, as illustrated by the maps (b) and (d) of figure~\ref{CORRAL},
we move from a highly localized signal around the pads (map (b)) to a broader response 
distributed along the corral rows (map (d)).

\section{Evolutionary optimization of metal nano-structures for maximum magnetic decay rate}

In order to demonstrate the versatility of our model, we couple our numerical framework to an evolutionary optimization (EO) algorithm.
EO tries to find optimum solutions to complex problems by mimicking the process of natural selection. 
Its principal idea is briefly depicted in figure~\ref{evolopt}a. 
Our approach to couple EO to numerical simulations is described in more detail in reference~\onlinecite{wiecha_evolutionary_2017}.
For technical information on the implementation and the used algorithm parameters, {\color{blue}see the SI}. In the {\color{blue}supporting information}, we also show an additional single- as well as a multi-objective evolutionary optimization problem, based on the decay-rate formalism.
In this section, we use the permittivity of gold\cite{johnson_optical_1972} to demonstrate that our formalism is not limited to dielectric materials. 
The optimization goal is to find a gold nano-structure which maximizes the ratio of magnetic over electric decay rate \(\Gamma_m/\Gamma_e\) at a fixed location (\(\mathbf{r}_0=(0,0,80)\)\,nm).
This is a particularly tricky scenario, because metals are known to have a far stronger response to electric dipole transitions than to magnetic ones.
We use the evolutionary algorithm to optimize the geometry of a planar structure composed of \(100\) gold pillars (each \(20\times 20\times 60\,\)nm\(^3\)), lying on a plane of \(1000\times 1000\,\)nm\(^2\) (see figure~\ref{evolopt}b).
We recall here that each subwavelength pillar does not support a direct magnetic response on its own.
To render the positioning easier, the possible locations on the plane lie on a discretized grid (steps of \(20\)\,nm). 
The structure is placed in vacuum and the wavelength is fixed at \(\lambda=500\,\)nm.
We evolve a population of 150 individuals (nano-structures) over \(2500\) generations. 
Each of the individuals is a parameter-set consisting of positions for the 100 gold pillars, hence describing one possible structure. 
We tested the convergence by running the same optimization several times, reproducibly yielding similar structures and values for the decay rate ratio.

The optimum structure found by the EO algorithm is shown in figure~\ref{evolopt}c. 
Mappings of the decay rate ratio as well as the electric and magnetic decay rates are shown in figure~\ref{evolopt}d-g.
Obviously, the algorithm succeeded in finding a gold nanostructure which significantly promotes magnetic decay at the target position (see Fig.~\ref{evolopt}d). 
This is particularly remarkable, because although gold structures easily provide very strong electric dipole decay rate enhancements, the magnetic LDOS is known to be usually very weak in metallic nanoparticles.\cite{albella_low-loss_2013}

Two effects are being exploited by the optimized structure: 
The first mechanism is the different confinement of the decay rates for electric and magnetic dipoles close to material. 
The electric decay rate enhancement in the proximity of the gold pillars is high, but confined to a very small volume around the material. 
The magnetic decay rate on the other hand is more loosely enhanced around the gold clusters, leading to regions in their vicinity where \(\Gamma_e\) is almost not affected, while \(\Gamma_m\) still shows significant enhancement (\textit{c.f.} figures~\ref{evolopt}e-f).
The second effect is a modulation of the decay rate inside a larger resonator due to interference, similar to the corral shown in figures~\ref{TAB-MAG} and~\ref{CORRAL}.
At \(\lambda=500\,\)nm, the above presented corral had a maximum of \(\Gamma_e\) in its center (see figure~\ref{CORRAL}b and~d).
In contrast to this, the evolutionary algorithm distributed a fraction of the material (outer, circular structure) such, that \(\Gamma_m\) is maximum in its center, which can be seen in figure~\ref{evolopt}g, where the decay rate has been calculated for the isolated outer structure.

We will conclude this section with some considerations on the convergence. 
One might wonder why the structure does not consist of perfect circles -- this would very likely result in even better performance. 
Concerning this question, we have to keep in mind that \((51 \times 51)! / (51 \times 51 - 100)! \approx 10^{341}\) possibilities exist to distribute the \(100\) gold pillars on the available positions on the plane.
Yet, the evolutionary algorithm did only evaluate \(2500\times 150 < 4\times 10^5\) different arrangements.
Therefore, the reason why the material is not distributed on perfect circles is the heuristic nature of the evolutionary optimization algorithm. 
The search for the best structure did simply not converge to the very optimum.
Comparing the optimized structure to an idealized version reveals, that the possible improvement in \(\Gamma_m/\Gamma_e\) is only in the order of \(\approx 1\)\% ({\color{blue}see also supporting information}).
We conclude that, despite the residual disorder in the geometry, the EO algorithm did converge very close to the ideal structure.
Hence EO is a promising approach to this kind of problems.

\noindent

\section{Conclusions and Perspectives}

In summary, we have developed a concise theoretical framework to describe the dynamics of light emission from magnetic dipoles located in complex nanostructured environments.
This method, based on mixed field-susceptibilities, provides analytical expressions of the decay rate in the case of very simple environments.
When the magnetic dipole is located close to nanostructures of arbitrary geometries, the computation of the MD decay rate involves the discretization of the nanostructure volume, the computation of a generalized propagator and finally the computation the decay rate from mixed field susceptibilities.
This versatile framework is well suited to describe the emission of light from emitters involving both electric and magnetic dipole transitions as well as nano-optical processes comprising confined electric and magnetic fields.
In addition, our framework is very flexible and can easily be extended. 
For instance nonlocality effects might be included by following the descriptions of Ref.~\onlinecite{girard_molecular_2015}.
Thanks to its computational simplicity, the method can also be employed within more complex numerical schemes. 
We demonstrated this possibility by coupling the magnetic decay rate calculation to an evolutionary optimization algorithm, which we employed to design a gold nanostructure for maximum contrast between magnetic and electric EM-LDOS.
We also applied our method to the decay rate close to complex dielectric nanostructures.
Our results suggest that it could be possible to identify the nature of the transition involved in the emission process (ED vs MD) from the variations of the decay rate in the vicinity of nanostructures.
Finally, nanostructures possessing a particularly high contrast regarding dipole orientations could be designed using our evolutionary optimization.

\begin{acknowledgments}

We thank Clément Majorel for his very useful contributions in the context of a Master's training.
This work was supported by Programme Investissements d'Avenir under the program ANR-11-IDEX-0002-02, reference ANR-10-LABX-0037-NEXT and by the computing facility center CALMIP of the University of Toulouse under grant P12167.

\end{acknowledgments}

\input{2017-decay-magnetic-dipole.bbl}

\end{document}

%% file: 2017-decay-magnetic-dipole.bbl
%

%% file: 2017-decay-magnetic-dipole.bbl
\begin{thebibliography}{50}%
\makeatletter
\providecommand \@ifxundefined [1]{%
 \@ifx{#1\undefined}
}%
\providecommand \@ifnum [1]{%
 \ifnum #1\expandafter \@firstoftwo
 \else \expandafter \@secondoftwo
 \fi
}%
\providecommand \@ifx [1]{%
 \ifx #1\expandafter \@firstoftwo
 \else \expandafter \@secondoftwo
 \fi
}%
\providecommand \natexlab [1]{#1}%
\providecommand \enquote  [1]{``#1''}%
\providecommand \bibnamefont  [1]{#1}%
\providecommand \bibfnamefont [1]{#1}%
\providecommand \citenamefont [1]{#1}%
\providecommand \href@noop [0]{\@secondoftwo}%
\providecommand \href [0]{\begingroup \@sanitize@url \@href}%
\providecommand \@href[1]{\@@startlink{#1}\@@href}%
\providecommand \@@href[1]{\endgroup#1\@@endlink}%
\providecommand \@sanitize@url [0]{\catcode `\\12\catcode `\$12\catcode
  `\&12\catcode `\#12\catcode `\^12\catcode `\_12\catcode `\%12\relax}%
\providecommand \@@startlink[1]{}%
\providecommand \@@endlink[0]{}%
\providecommand \url  [0]{\begingroup\@sanitize@url \@url }%
\providecommand \@url [1]{\endgroup\@href {#1}{\urlprefix }}%
\providecommand \urlprefix  [0]{URL }%
\providecommand \Eprint [0]{\href }%
\providecommand \doibase [0]{http://dx.doi.org/}%
\providecommand \selectlanguage [0]{\@gobble}%
\providecommand \bibinfo  [0]{\@secondoftwo}%
\providecommand \bibfield  [0]{\@secondoftwo}%
\providecommand \translation [1]{[#1]}%
\providecommand \BibitemOpen [0]{}%
\providecommand \bibitemStop [0]{}%
\providecommand \bibitemNoStop [0]{.\EOS\space}%
\providecommand \EOS [0]{\spacefactor3000\relax}%
\providecommand \BibitemShut  [1]{\csname bibitem#1\endcsname}%
\let\auto@bib@innerbib\@empty
\bibitem [{\citenamefont {Novotny}\ and\ \citenamefont
  {Hecht}(2006)}]{novotny_principles_2006}%
  \BibitemOpen
  \bibfield  {author} {\bibinfo {author} {\bibfnamefont {L.}~\bibnamefont
  {Novotny}}\ and\ \bibinfo {author} {\bibfnamefont {B.}~\bibnamefont
  {Hecht}},\ }\href@noop {} {\emph {\bibinfo {title} {Principles of
  Nano-Optics}}}\ (\bibinfo  {publisher} {{Cambridge University Press}},\
  \bibinfo {address} {Cambridge ; New York},\ \bibinfo {year}
  {2006})\BibitemShut {NoStop}%
\bibitem [{\citenamefont {Anger}\ \emph {et~al.}(2006)\citenamefont {Anger},
  \citenamefont {Bharadwaj},\ and\ \citenamefont
  {Novotny}}]{anger_enhancement_2006}%
  \BibitemOpen
  \bibfield  {author} {\bibinfo {author} {\bibfnamefont {P.}~\bibnamefont
  {Anger}}, \bibinfo {author} {\bibfnamefont {P.}~\bibnamefont {Bharadwaj}}, \
  and\ \bibinfo {author} {\bibfnamefont {L.}~\bibnamefont {Novotny}},\ }\href
  {\doibase 10.1103/PhysRevLett.96.113002} {\bibfield  {journal} {\bibinfo
  {journal} {Physical Review Letters}\ }\textbf {\bibinfo {volume} {96}},\
  \bibinfo {pages} {113002} (\bibinfo {year} {2006})}\BibitemShut {NoStop}%
\bibitem [{\citenamefont {K{\"u}hn}\ \emph {et~al.}(2006)\citenamefont
  {K{\"u}hn}, \citenamefont {H{\aa}kanson}, \citenamefont {Rogobete},\ and\
  \citenamefont {Sandoghdar}}]{kuhn_enhancement_2006}%
  \BibitemOpen
  \bibfield  {author} {\bibinfo {author} {\bibfnamefont {S.}~\bibnamefont
  {K{\"u}hn}}, \bibinfo {author} {\bibfnamefont {U.}~\bibnamefont
  {H{\aa}kanson}}, \bibinfo {author} {\bibfnamefont {L.}~\bibnamefont
  {Rogobete}}, \ and\ \bibinfo {author} {\bibfnamefont {V.}~\bibnamefont
  {Sandoghdar}},\ }\href {\doibase 10.1103/PhysRevLett.97.017402} {\bibfield
  {journal} {\bibinfo  {journal} {Physical Review Letters}\ }\textbf {\bibinfo
  {volume} {97}},\ \bibinfo {pages} {017402} (\bibinfo {year}
  {2006})}\BibitemShut {NoStop}%
\bibitem [{\citenamefont {Kinkhabwala}\ \emph {et~al.}(2009)\citenamefont
  {Kinkhabwala}, \citenamefont {Yu}, \citenamefont {Fan}, \citenamefont
  {Avlasevich}, \citenamefont {M{\"u}llen},\ and\ \citenamefont
  {Moerner}}]{kinkhabwala_large_2009}%
  \BibitemOpen
  \bibfield  {author} {\bibinfo {author} {\bibfnamefont {A.}~\bibnamefont
  {Kinkhabwala}}, \bibinfo {author} {\bibfnamefont {Z.}~\bibnamefont {Yu}},
  \bibinfo {author} {\bibfnamefont {S.}~\bibnamefont {Fan}}, \bibinfo {author}
  {\bibfnamefont {Y.}~\bibnamefont {Avlasevich}}, \bibinfo {author}
  {\bibfnamefont {K.}~\bibnamefont {M{\"u}llen}}, \ and\ \bibinfo {author}
  {\bibfnamefont {W.~E.}\ \bibnamefont {Moerner}},\ }\href {\doibase
  10.1038/nphoton.2009.187} {\bibfield  {journal} {\bibinfo  {journal} {Nature
  Photonics}\ }\textbf {\bibinfo {volume} {3}},\ \bibinfo {pages} {654}
  (\bibinfo {year} {2009})}\BibitemShut {NoStop}%
\bibitem [{\citenamefont {Curto}\ \emph {et~al.}(2010)\citenamefont {Curto},
  \citenamefont {Volpe}, \citenamefont {Taminiau}, \citenamefont {Kreuzer},
  \citenamefont {Quidant},\ and\ \citenamefont {van
  Hulst}}]{curto_unidirectional_2010}%
  \BibitemOpen
  \bibfield  {author} {\bibinfo {author} {\bibfnamefont {A.~G.}\ \bibnamefont
  {Curto}}, \bibinfo {author} {\bibfnamefont {G.}~\bibnamefont {Volpe}},
  \bibinfo {author} {\bibfnamefont {T.~H.}\ \bibnamefont {Taminiau}}, \bibinfo
  {author} {\bibfnamefont {M.~P.}\ \bibnamefont {Kreuzer}}, \bibinfo {author}
  {\bibfnamefont {R.}~\bibnamefont {Quidant}}, \ and\ \bibinfo {author}
  {\bibfnamefont {N.~F.}\ \bibnamefont {van Hulst}},\ }\href {\doibase
  10.1126/science.1191922} {\bibfield  {journal} {\bibinfo  {journal}
  {Science}\ }\textbf {\bibinfo {volume} {329}},\ \bibinfo {pages} {930}
  (\bibinfo {year} {2010})}\BibitemShut {NoStop}%
\bibitem [{\citenamefont {Biagioni}\ \emph {et~al.}(2012)\citenamefont
  {Biagioni}, \citenamefont {Huang},\ and\ \citenamefont
  {Hecht}}]{biagioni_nanoantennas_2012}%
  \BibitemOpen
  \bibfield  {author} {\bibinfo {author} {\bibfnamefont {P.}~\bibnamefont
  {Biagioni}}, \bibinfo {author} {\bibfnamefont {J.-S.}\ \bibnamefont {Huang}},
  \ and\ \bibinfo {author} {\bibfnamefont {B.}~\bibnamefont {Hecht}},\
  }\href@noop {} {\bibfield  {journal} {\bibinfo  {journal} {Reports on
  Progress in Physics}\ }\textbf {\bibinfo {volume} {75}},\ \bibinfo {pages}
  {024402} (\bibinfo {year} {2012})}\BibitemShut {NoStop}%
\bibitem [{\citenamefont {Novotny}\ and\ \citenamefont {{van
  Hulst}}(2011)}]{novotny_antennas_2011}%
  \BibitemOpen
  \bibfield  {author} {\bibinfo {author} {\bibfnamefont {L.}~\bibnamefont
  {Novotny}}\ and\ \bibinfo {author} {\bibfnamefont {N.}~\bibnamefont {{van
  Hulst}}},\ }\href {\doibase 10.1038/nphoton.2010.237} {\bibfield  {journal}
  {\bibinfo  {journal} {Nature Photonics}\ }\textbf {\bibinfo {volume} {5}},\
  \bibinfo {pages} {83} (\bibinfo {year} {2011})}\BibitemShut {NoStop}%
\bibitem [{\citenamefont {Klimov}\ \emph {et~al.}(1996)\citenamefont {Klimov},
  \citenamefont {Ducloy},\ and\ \citenamefont
  {Letokhov}}]{klimov_radiative_1996}%
  \BibitemOpen
  \bibfield  {author} {\bibinfo {author} {\bibfnamefont {V.~V.}\ \bibnamefont
  {Klimov}}, \bibinfo {author} {\bibfnamefont {M.}~\bibnamefont {Ducloy}}, \
  and\ \bibinfo {author} {\bibfnamefont {V.~S.}\ \bibnamefont {Letokhov}},\
  }\href {\doibase 10.1080/09500349608232884} {\bibfield  {journal} {\bibinfo
  {journal} {Journal of Modern Optics}\ }\textbf {\bibinfo {volume} {43}},\
  \bibinfo {pages} {2251} (\bibinfo {year} {1996})}\BibitemShut {NoStop}%
\bibitem [{\citenamefont {Giessen}\ and\ \citenamefont
  {Vogelgesang}(2009)}]{giessen_glimpsing_2009}%
  \BibitemOpen
  \bibfield  {author} {\bibinfo {author} {\bibfnamefont {H.}~\bibnamefont
  {Giessen}}\ and\ \bibinfo {author} {\bibfnamefont {R.}~\bibnamefont
  {Vogelgesang}},\ }\href {\doibase 10.1126/science.1181552} {\bibfield
  {journal} {\bibinfo  {journal} {Science}\ }\textbf {\bibinfo {volume}
  {326}},\ \bibinfo {pages} {529} (\bibinfo {year} {2009})}\BibitemShut
  {NoStop}%
\bibitem [{\citenamefont {Aigouy}\ \emph {et~al.}(2014)\citenamefont {Aigouy},
  \citenamefont {Caz{\'e}}, \citenamefont {Gredin}, \citenamefont {Mortier},\
  and\ \citenamefont {Carminati}}]{aigouy_mapping_2014}%
  \BibitemOpen
  \bibfield  {author} {\bibinfo {author} {\bibfnamefont {L.}~\bibnamefont
  {Aigouy}}, \bibinfo {author} {\bibfnamefont {A.}~\bibnamefont {Caz{\'e}}},
  \bibinfo {author} {\bibfnamefont {P.}~\bibnamefont {Gredin}}, \bibinfo
  {author} {\bibfnamefont {M.}~\bibnamefont {Mortier}}, \ and\ \bibinfo
  {author} {\bibfnamefont {R.}~\bibnamefont {Carminati}},\ }\href {\doibase
  10.1103/PhysRevLett.113.076101} {\bibfield  {journal} {\bibinfo  {journal}
  {Physical Review Letters}\ }\textbf {\bibinfo {volume} {113}},\ \bibinfo
  {pages} {076101} (\bibinfo {year} {2014})}\BibitemShut {NoStop}%
\bibitem [{\citenamefont {Choi}\ \emph {et~al.}(2016)\citenamefont {Choi},
  \citenamefont {Iwanaga}, \citenamefont {Sugimoto}, \citenamefont {Sakoda},\
  and\ \citenamefont {Miyazaki}}]{choi_selective_2016}%
  \BibitemOpen
  \bibfield  {author} {\bibinfo {author} {\bibfnamefont {B.}~\bibnamefont
  {Choi}}, \bibinfo {author} {\bibfnamefont {M.}~\bibnamefont {Iwanaga}},
  \bibinfo {author} {\bibfnamefont {Y.}~\bibnamefont {Sugimoto}}, \bibinfo
  {author} {\bibfnamefont {K.}~\bibnamefont {Sakoda}}, \ and\ \bibinfo {author}
  {\bibfnamefont {H.~T.}\ \bibnamefont {Miyazaki}},\ }\href {\doibase
  10.1021/acs.nanolett.6b02200} {\bibfield  {journal} {\bibinfo  {journal}
  {Nano Letters}\ }\textbf {\bibinfo {volume} {16}},\ \bibinfo {pages} {5191}
  (\bibinfo {year} {2016})}\BibitemShut {NoStop}%
\bibitem [{\citenamefont {Karaveli}\ and\ \citenamefont
  {Zia}(2011)}]{karaveli_spectral_2011}%
  \BibitemOpen
  \bibfield  {author} {\bibinfo {author} {\bibfnamefont {S.}~\bibnamefont
  {Karaveli}}\ and\ \bibinfo {author} {\bibfnamefont {R.}~\bibnamefont {Zia}},\
  }\href {\doibase 10.1103/PhysRevLett.106.193004} {\bibfield  {journal}
  {\bibinfo  {journal} {Physical Review Letters}\ }\textbf {\bibinfo {volume}
  {106}},\ \bibinfo {pages} {193004} (\bibinfo {year} {2011})}\BibitemShut
  {NoStop}%
\bibitem [{\citenamefont {Carminati}\ \emph {et~al.}(2015)\citenamefont
  {Carminati}, \citenamefont {Caz{\'e}}, \citenamefont {Cao}, \citenamefont
  {Peragut}, \citenamefont {Krachmalnicoff}, \citenamefont {Pierrat},\ and\
  \citenamefont {De~Wilde}}]{carminati_electromagnetic_2015}%
  \BibitemOpen
  \bibfield  {author} {\bibinfo {author} {\bibfnamefont {R.}~\bibnamefont
  {Carminati}}, \bibinfo {author} {\bibfnamefont {A.}~\bibnamefont {Caz{\'e}}},
  \bibinfo {author} {\bibfnamefont {D.}~\bibnamefont {Cao}}, \bibinfo {author}
  {\bibfnamefont {F.}~\bibnamefont {Peragut}}, \bibinfo {author} {\bibfnamefont
  {V.}~\bibnamefont {Krachmalnicoff}}, \bibinfo {author} {\bibfnamefont
  {R.}~\bibnamefont {Pierrat}}, \ and\ \bibinfo {author} {\bibfnamefont
  {Y.}~\bibnamefont {De~Wilde}},\ }\href {\doibase
  10.1016/j.surfrep.2014.11.001} {\bibfield  {journal} {\bibinfo  {journal}
  {Surface Science Reports}\ }\textbf {\bibinfo {volume} {70}},\ \bibinfo
  {pages} {1} (\bibinfo {year} {2015})}\BibitemShut {NoStop}%
\bibitem [{\citenamefont {Baranov}\ \emph {et~al.}(2017)\citenamefont
  {Baranov}, \citenamefont {Savelev}, \citenamefont {Li}, \citenamefont
  {Krasnok},\ and\ \citenamefont {Al{\`u}}}]{baranov_modifying_2017}%
  \BibitemOpen
  \bibfield  {author} {\bibinfo {author} {\bibfnamefont {D.~G.}\ \bibnamefont
  {Baranov}}, \bibinfo {author} {\bibfnamefont {R.~S.}\ \bibnamefont
  {Savelev}}, \bibinfo {author} {\bibfnamefont {S.~V.}\ \bibnamefont {Li}},
  \bibinfo {author} {\bibfnamefont {A.~E.}\ \bibnamefont {Krasnok}}, \ and\
  \bibinfo {author} {\bibfnamefont {A.}~\bibnamefont {Al{\`u}}},\ }\href
  {\doibase 10.1002/lpor.201600268} {\bibfield  {journal} {\bibinfo  {journal}
  {Laser \& Photonics Reviews}\ }\textbf {\bibinfo {volume} {11}},\ \bibinfo
  {pages} {1600268} (\bibinfo {year} {2017})}\BibitemShut {NoStop}%
\bibitem [{\citenamefont {Cuche}\ \emph {et~al.}(2017)\citenamefont {Cuche},
  \citenamefont {Berthel}, \citenamefont {Kumar}, \citenamefont {{Colas des
  Francs}}, \citenamefont {Huant}, \citenamefont {Dujardin}, \citenamefont
  {Girard},\ and\ \citenamefont {Drezet}}]{cuche_near-field_2017}%
  \BibitemOpen
  \bibfield  {author} {\bibinfo {author} {\bibfnamefont {A.}~\bibnamefont
  {Cuche}}, \bibinfo {author} {\bibfnamefont {M.}~\bibnamefont {Berthel}},
  \bibinfo {author} {\bibfnamefont {U.}~\bibnamefont {Kumar}}, \bibinfo
  {author} {\bibfnamefont {G.}~\bibnamefont {{Colas des Francs}}}, \bibinfo
  {author} {\bibfnamefont {S.}~\bibnamefont {Huant}}, \bibinfo {author}
  {\bibfnamefont {E.}~\bibnamefont {Dujardin}}, \bibinfo {author}
  {\bibfnamefont {C.}~\bibnamefont {Girard}}, \ and\ \bibinfo {author}
  {\bibfnamefont {A.}~\bibnamefont {Drezet}},\ }\href {\doibase
  10.1103/PhysRevB.95.121402} {\bibfield  {journal} {\bibinfo  {journal}
  {Physical Review B}\ }\textbf {\bibinfo {volume} {95}},\ \bibinfo {pages}
  {121402} (\bibinfo {year} {2017})}\BibitemShut {NoStop}%
\bibitem [{\citenamefont {Purcell}\ \emph {et~al.}(1946)\citenamefont
  {Purcell}, \citenamefont {Torrey},\ and\ \citenamefont
  {Pound}}]{purcell_resonance_1946}%
  \BibitemOpen
  \bibfield  {author} {\bibinfo {author} {\bibfnamefont {E.~M.}\ \bibnamefont
  {Purcell}}, \bibinfo {author} {\bibfnamefont {H.~C.}\ \bibnamefont {Torrey}},
  \ and\ \bibinfo {author} {\bibfnamefont {R.~V.}\ \bibnamefont {Pound}},\
  }\href {\doibase 10.1103/PhysRev.69.37} {\bibfield  {journal} {\bibinfo
  {journal} {Physical Review}\ }\textbf {\bibinfo {volume} {69}},\ \bibinfo
  {pages} {37} (\bibinfo {year} {1946})}\BibitemShut {NoStop}%
\bibitem [{\citenamefont {Chicanne}\ \emph {et~al.}(2002)\citenamefont
  {Chicanne}, \citenamefont {David}, \citenamefont {Quidant}, \citenamefont
  {Weeber}, \citenamefont {Lacroute}, \citenamefont {Bourillot}, \citenamefont
  {Dereux}, \citenamefont {{Colas des Francs}},\ and\ \citenamefont
  {Girard}}]{chicanne_imaging_2002}%
  \BibitemOpen
  \bibfield  {author} {\bibinfo {author} {\bibfnamefont {C.}~\bibnamefont
  {Chicanne}}, \bibinfo {author} {\bibfnamefont {T.}~\bibnamefont {David}},
  \bibinfo {author} {\bibfnamefont {R.}~\bibnamefont {Quidant}}, \bibinfo
  {author} {\bibfnamefont {J.~C.}\ \bibnamefont {Weeber}}, \bibinfo {author}
  {\bibfnamefont {Y.}~\bibnamefont {Lacroute}}, \bibinfo {author}
  {\bibfnamefont {E.}~\bibnamefont {Bourillot}}, \bibinfo {author}
  {\bibfnamefont {A.}~\bibnamefont {Dereux}}, \bibinfo {author} {\bibfnamefont
  {G.}~\bibnamefont {{Colas des Francs}}}, \ and\ \bibinfo {author}
  {\bibfnamefont {C.}~\bibnamefont {Girard}},\ }\href {\doibase
  10.1103/PhysRevLett.88.097402} {\bibfield  {journal} {\bibinfo  {journal}
  {Physical Review Letters}\ }\textbf {\bibinfo {volume} {88}},\ \bibinfo
  {pages} {097402} (\bibinfo {year} {2002})}\BibitemShut {NoStop}%
\bibitem [{\citenamefont {Rolly}\ \emph {et~al.}(2012)\citenamefont {Rolly},
  \citenamefont {Bebey}, \citenamefont {Bidault}, \citenamefont {Stout},\ and\
  \citenamefont {Bonod}}]{rolly_promoting_2012}%
  \BibitemOpen
  \bibfield  {author} {\bibinfo {author} {\bibfnamefont {B.}~\bibnamefont
  {Rolly}}, \bibinfo {author} {\bibfnamefont {B.}~\bibnamefont {Bebey}},
  \bibinfo {author} {\bibfnamefont {S.}~\bibnamefont {Bidault}}, \bibinfo
  {author} {\bibfnamefont {B.}~\bibnamefont {Stout}}, \ and\ \bibinfo {author}
  {\bibfnamefont {N.}~\bibnamefont {Bonod}},\ }\href {\doibase
  10.1103/PhysRevB.85.245432} {\bibfield  {journal} {\bibinfo  {journal}
  {Physical Review B}\ }\textbf {\bibinfo {volume} {85}},\ \bibinfo {pages}
  {245432} (\bibinfo {year} {2012})}\BibitemShut {NoStop}%
\bibitem [{\citenamefont {Chigrin}\ \emph {et~al.}(2016)\citenamefont
  {Chigrin}, \citenamefont {Kumar}, \citenamefont {Cuma},\ and\ \citenamefont
  {{von Plessen}}}]{chigrin_emission_2016}%
  \BibitemOpen
  \bibfield  {author} {\bibinfo {author} {\bibfnamefont {D.~N.}\ \bibnamefont
  {Chigrin}}, \bibinfo {author} {\bibfnamefont {D.}~\bibnamefont {Kumar}},
  \bibinfo {author} {\bibfnamefont {D.}~\bibnamefont {Cuma}}, \ and\ \bibinfo
  {author} {\bibfnamefont {G.}~\bibnamefont {{von Plessen}}},\ }\href {\doibase
  10.1021/acsphotonics.5b00397} {\bibfield  {journal} {\bibinfo  {journal} {ACS
  Photonics}\ }\textbf {\bibinfo {volume} {3}},\ \bibinfo {pages} {27}
  (\bibinfo {year} {2016})}\BibitemShut {NoStop}%
\bibitem [{\citenamefont {Chew}(1979)}]{chew_fluorescent_1979}%
  \BibitemOpen
  \bibfield  {author} {\bibinfo {author} {\bibfnamefont {H.}~\bibnamefont
  {Chew}},\ }\href {\doibase 10.1103/PhysRevA.19.2137} {\bibfield  {journal}
  {\bibinfo  {journal} {Physical Review A}\ }\textbf {\bibinfo {volume} {19}},\
  \bibinfo {pages} {2137} (\bibinfo {year} {1979})}\BibitemShut {NoStop}%
\bibitem [{\citenamefont {Klimov}\ and\ \citenamefont
  {Letokhov}(2005)}]{klimov_electric_2005}%
  \BibitemOpen
  \bibfield  {author} {\bibinfo {author} {\bibfnamefont {V.~V.}\ \bibnamefont
  {Klimov}}\ and\ \bibinfo {author} {\bibfnamefont {V.~S.}\ \bibnamefont
  {Letokhov}},\ }\href@noop {} {\bibfield  {journal} {\bibinfo  {journal}
  {Laser physics}\ }\textbf {\bibinfo {volume} {15}},\ \bibinfo {pages} {61}
  (\bibinfo {year} {2005})}\BibitemShut {NoStop}%
\bibitem [{\citenamefont {Schmidt}\ \emph {et~al.}(2012)\citenamefont
  {Schmidt}, \citenamefont {Esteban}, \citenamefont {S{\'a}enz}, \citenamefont
  {Su{\'a}rez-Lacalle}, \citenamefont {Mackowski},\ and\ \citenamefont
  {Aizpurua}}]{schmidt_dielectric_2012}%
  \BibitemOpen
  \bibfield  {author} {\bibinfo {author} {\bibfnamefont {M.~K.}\ \bibnamefont
  {Schmidt}}, \bibinfo {author} {\bibfnamefont {R.}~\bibnamefont {Esteban}},
  \bibinfo {author} {\bibfnamefont {J.~J.}\ \bibnamefont {S{\'a}enz}}, \bibinfo
  {author} {\bibfnamefont {I.}~\bibnamefont {Su{\'a}rez-Lacalle}}, \bibinfo
  {author} {\bibfnamefont {S.}~\bibnamefont {Mackowski}}, \ and\ \bibinfo
  {author} {\bibfnamefont {J.}~\bibnamefont {Aizpurua}},\ }\href {\doibase
  10.1364/OE.20.013636} {\bibfield  {journal} {\bibinfo  {journal} {Optics
  Express}\ }\textbf {\bibinfo {volume} {20}},\ \bibinfo {pages} {13636}
  (\bibinfo {year} {2012})}\BibitemShut {NoStop}%
\bibitem [{\citenamefont {Stout}\ \emph {et~al.}(2011)\citenamefont {Stout},
  \citenamefont {Devilez}, \citenamefont {Rolly},\ and\ \citenamefont
  {Bonod}}]{stout_multipole_2011}%
  \BibitemOpen
  \bibfield  {author} {\bibinfo {author} {\bibfnamefont {B.}~\bibnamefont
  {Stout}}, \bibinfo {author} {\bibfnamefont {A.}~\bibnamefont {Devilez}},
  \bibinfo {author} {\bibfnamefont {B.}~\bibnamefont {Rolly}}, \ and\ \bibinfo
  {author} {\bibfnamefont {N.}~\bibnamefont {Bonod}},\ }\href {\doibase
  10.1364/JOSAB.28.001213} {\bibfield  {journal} {\bibinfo  {journal} {JOSA B}\
  }\textbf {\bibinfo {volume} {28}},\ \bibinfo {pages} {1213} (\bibinfo {year}
  {2011})}\BibitemShut {NoStop}%
\bibitem [{\citenamefont {Feng}\ \emph {et~al.}(2011)\citenamefont {Feng},
  \citenamefont {Zhou}, \citenamefont {Liu},\ and\ \citenamefont
  {Li}}]{feng_controlling_2011}%
  \BibitemOpen
  \bibfield  {author} {\bibinfo {author} {\bibfnamefont {T.}~\bibnamefont
  {Feng}}, \bibinfo {author} {\bibfnamefont {Y.}~\bibnamefont {Zhou}}, \bibinfo
  {author} {\bibfnamefont {D.}~\bibnamefont {Liu}}, \ and\ \bibinfo {author}
  {\bibfnamefont {J.}~\bibnamefont {Li}},\ }\href {\doibase
  10.1364/OL.36.002369} {\bibfield  {journal} {\bibinfo  {journal} {Optics
  Letters}\ }\textbf {\bibinfo {volume} {36}},\ \bibinfo {pages} {2369}
  (\bibinfo {year} {2011})}\BibitemShut {NoStop}%
\bibitem [{\citenamefont {Hein}\ and\ \citenamefont
  {Giessen}(2013)}]{hein_tailoring_2013}%
  \BibitemOpen
  \bibfield  {author} {\bibinfo {author} {\bibfnamefont {S.~M.}\ \bibnamefont
  {Hein}}\ and\ \bibinfo {author} {\bibfnamefont {H.}~\bibnamefont {Giessen}},\
  }\href {\doibase 10.1103/PhysRevLett.111.026803} {\bibfield  {journal}
  {\bibinfo  {journal} {Physical Review Letters}\ }\textbf {\bibinfo {volume}
  {111}},\ \bibinfo {pages} {026803} (\bibinfo {year} {2013})}\BibitemShut
  {NoStop}%
\bibitem [{\citenamefont {Albella}\ \emph {et~al.}(2013)\citenamefont
  {Albella}, \citenamefont {Poyli}, \citenamefont {Schmidt}, \citenamefont
  {Maier}, \citenamefont {Moreno}, \citenamefont {S{\'a}enz},\ and\
  \citenamefont {Aizpurua}}]{albella_low-loss_2013}%
  \BibitemOpen
  \bibfield  {author} {\bibinfo {author} {\bibfnamefont {P.}~\bibnamefont
  {Albella}}, \bibinfo {author} {\bibfnamefont {M.~A.}\ \bibnamefont {Poyli}},
  \bibinfo {author} {\bibfnamefont {M.~K.}\ \bibnamefont {Schmidt}}, \bibinfo
  {author} {\bibfnamefont {S.~A.}\ \bibnamefont {Maier}}, \bibinfo {author}
  {\bibfnamefont {F.}~\bibnamefont {Moreno}}, \bibinfo {author} {\bibfnamefont
  {J.~J.}\ \bibnamefont {S{\'a}enz}}, \ and\ \bibinfo {author} {\bibfnamefont
  {J.}~\bibnamefont {Aizpurua}},\ }\href {\doibase 10.1021/jp4027018}
  {\bibfield  {journal} {\bibinfo  {journal} {The Journal of Physical Chemistry
  C}\ }\textbf {\bibinfo {volume} {117}},\ \bibinfo {pages} {13573} (\bibinfo
  {year} {2013})}\BibitemShut {NoStop}%
\bibitem [{\citenamefont {Mivelle}\ \emph {et~al.}(2015)\citenamefont
  {Mivelle}, \citenamefont {Grosjean}, \citenamefont {Burr}, \citenamefont
  {Fischer},\ and\ \citenamefont {Garcia-Parajo}}]{mivelle_strong_2015}%
  \BibitemOpen
  \bibfield  {author} {\bibinfo {author} {\bibfnamefont {M.}~\bibnamefont
  {Mivelle}}, \bibinfo {author} {\bibfnamefont {T.}~\bibnamefont {Grosjean}},
  \bibinfo {author} {\bibfnamefont {G.~W.}\ \bibnamefont {Burr}}, \bibinfo
  {author} {\bibfnamefont {U.~C.}\ \bibnamefont {Fischer}}, \ and\ \bibinfo
  {author} {\bibfnamefont {M.~F.}\ \bibnamefont {Garcia-Parajo}},\ }\href
  {\doibase 10.1021/acsphotonics.5b00128} {\bibfield  {journal} {\bibinfo
  {journal} {ACS Photonics}\ }\textbf {\bibinfo {volume} {2}},\ \bibinfo
  {pages} {1071} (\bibinfo {year} {2015})}\BibitemShut {NoStop}%
\bibitem [{\citenamefont {Feng}\ \emph {et~al.}(2016)\citenamefont {Feng},
  \citenamefont {Xu}, \citenamefont {Liang},\ and\ \citenamefont
  {Zhang}}]{feng_all-dielectric_2016}%
  \BibitemOpen
  \bibfield  {author} {\bibinfo {author} {\bibfnamefont {T.}~\bibnamefont
  {Feng}}, \bibinfo {author} {\bibfnamefont {Y.}~\bibnamefont {Xu}}, \bibinfo
  {author} {\bibfnamefont {Z.}~\bibnamefont {Liang}}, \ and\ \bibinfo {author}
  {\bibfnamefont {W.}~\bibnamefont {Zhang}},\ }\href {\doibase
  10.1364/OL.41.005011} {\bibfield  {journal} {\bibinfo  {journal} {Optics
  Letters}\ }\textbf {\bibinfo {volume} {41}},\ \bibinfo {pages} {5011}
  (\bibinfo {year} {2016})}\BibitemShut {NoStop}%
\bibitem [{\citenamefont {Burresi}\ \emph {et~al.}(2009)\citenamefont
  {Burresi}, \citenamefont {van Oosten}, \citenamefont {Kampfrath},
  \citenamefont {Schoenmaker}, \citenamefont {Heideman}, \citenamefont
  {Leinse},\ and\ \citenamefont {Kuipers}}]{burresi_probing_2009}%
  \BibitemOpen
  \bibfield  {author} {\bibinfo {author} {\bibfnamefont {M.}~\bibnamefont
  {Burresi}}, \bibinfo {author} {\bibfnamefont {D.}~\bibnamefont {van Oosten}},
  \bibinfo {author} {\bibfnamefont {T.}~\bibnamefont {Kampfrath}}, \bibinfo
  {author} {\bibfnamefont {H.}~\bibnamefont {Schoenmaker}}, \bibinfo {author}
  {\bibfnamefont {R.}~\bibnamefont {Heideman}}, \bibinfo {author}
  {\bibfnamefont {A.}~\bibnamefont {Leinse}}, \ and\ \bibinfo {author}
  {\bibfnamefont {L.}~\bibnamefont {Kuipers}},\ }\href {\doibase
  10.1126/science.1177096} {\bibfield  {journal} {\bibinfo  {journal}
  {Science}\ }\textbf {\bibinfo {volume} {326}},\ \bibinfo {pages} {550}
  (\bibinfo {year} {2009})}\BibitemShut {NoStop}%
\bibitem [{\citenamefont {Devaux}\ \emph
  {et~al.}(2000{\natexlab{a}})\citenamefont {Devaux}, \citenamefont {Dereux},
  \citenamefont {Bourillot}, \citenamefont {Weeber}, \citenamefont {Lacroute},
  \citenamefont {Goudonnet},\ and\ \citenamefont
  {Girard}}]{devaux_detection_2000}%
  \BibitemOpen
  \bibfield  {author} {\bibinfo {author} {\bibfnamefont {E.}~\bibnamefont
  {Devaux}}, \bibinfo {author} {\bibfnamefont {A.}~\bibnamefont {Dereux}},
  \bibinfo {author} {\bibfnamefont {E.}~\bibnamefont {Bourillot}}, \bibinfo
  {author} {\bibfnamefont {J.-C.}\ \bibnamefont {Weeber}}, \bibinfo {author}
  {\bibfnamefont {Y.}~\bibnamefont {Lacroute}}, \bibinfo {author}
  {\bibfnamefont {J.-P.}\ \bibnamefont {Goudonnet}}, \ and\ \bibinfo {author}
  {\bibfnamefont {C.}~\bibnamefont {Girard}},\ }\href {\doibase
  10.1016/S0169-4332(00)00349-4} {\bibfield  {journal} {\bibinfo  {journal}
  {Applied Surface Science}\ }\textbf {\bibinfo {volume} {164}},\ \bibinfo
  {pages} {124} (\bibinfo {year} {2000}{\natexlab{a}})}\BibitemShut {NoStop}%
\bibitem [{\citenamefont {Devaux}\ \emph
  {et~al.}(2000{\natexlab{b}})\citenamefont {Devaux}, \citenamefont {Dereux},
  \citenamefont {Bourillot}, \citenamefont {Weeber}, \citenamefont {Lacroute},
  \citenamefont {Goudonnet},\ and\ \citenamefont {Girard}}]{devaux_local_2000}%
  \BibitemOpen
  \bibfield  {author} {\bibinfo {author} {\bibfnamefont {E.}~\bibnamefont
  {Devaux}}, \bibinfo {author} {\bibfnamefont {A.}~\bibnamefont {Dereux}},
  \bibinfo {author} {\bibfnamefont {E.}~\bibnamefont {Bourillot}}, \bibinfo
  {author} {\bibfnamefont {J.-C.}\ \bibnamefont {Weeber}}, \bibinfo {author}
  {\bibfnamefont {Y.}~\bibnamefont {Lacroute}}, \bibinfo {author}
  {\bibfnamefont {J.-P.}\ \bibnamefont {Goudonnet}}, \ and\ \bibinfo {author}
  {\bibfnamefont {C.}~\bibnamefont {Girard}},\ }\href {\doibase
  10.1103/PhysRevB.62.10504} {\bibfield  {journal} {\bibinfo  {journal}
  {Physical Review B}\ }\textbf {\bibinfo {volume} {62}},\ \bibinfo {pages}
  {10504} (\bibinfo {year} {2000}{\natexlab{b}})}\BibitemShut {NoStop}%
\bibitem [{\citenamefont {Girard}\ \emph {et~al.}(1997)\citenamefont {Girard},
  \citenamefont {Weeber}, \citenamefont {Dereux}, \citenamefont {Martin},\ and\
  \citenamefont {Goudonnet}}]{girard_optical_1997}%
  \BibitemOpen
  \bibfield  {author} {\bibinfo {author} {\bibfnamefont {C.}~\bibnamefont
  {Girard}}, \bibinfo {author} {\bibfnamefont {J.-C.}\ \bibnamefont {Weeber}},
  \bibinfo {author} {\bibfnamefont {A.}~\bibnamefont {Dereux}}, \bibinfo
  {author} {\bibfnamefont {O.~J.~F.}\ \bibnamefont {Martin}}, \ and\ \bibinfo
  {author} {\bibfnamefont {J.-P.}\ \bibnamefont {Goudonnet}},\ }\href {\doibase
  10.1103/PhysRevB.55.16487} {\bibfield  {journal} {\bibinfo  {journal}
  {Physical Review B}\ }\textbf {\bibinfo {volume} {55}},\ \bibinfo {pages}
  {16487} (\bibinfo {year} {1997})}\BibitemShut {NoStop}%
\bibitem [{\citenamefont {Schr{\"o}ter}(2003)}]{schroter_modelling_2003}%
  \BibitemOpen
  \bibfield  {author} {\bibinfo {author} {\bibfnamefont {U.}~\bibnamefont
  {Schr{\"o}ter}},\ }\href {\doibase 10.1140/epjb/e2003-00170-y} {\bibfield
  {journal} {\bibinfo  {journal} {The European Physical Journal B}\ }\textbf
  {\bibinfo {volume} {33}},\ \bibinfo {pages} {297} (\bibinfo {year}
  {2003})}\BibitemShut {NoStop}%
\bibitem [{\citenamefont {Martin}\ \emph {et~al.}(1995)\citenamefont {Martin},
  \citenamefont {Girard},\ and\ \citenamefont
  {Dereux}}]{martin_generalized_1995}%
  \BibitemOpen
  \bibfield  {author} {\bibinfo {author} {\bibfnamefont {O.~J.~F.}\
  \bibnamefont {Martin}}, \bibinfo {author} {\bibfnamefont {C.}~\bibnamefont
  {Girard}}, \ and\ \bibinfo {author} {\bibfnamefont {A.}~\bibnamefont
  {Dereux}},\ }\href {\doibase 10.1103/PhysRevLett.74.526} {\bibfield
  {journal} {\bibinfo  {journal} {Physical Review Letters}\ }\textbf {\bibinfo
  {volume} {74}},\ \bibinfo {pages} {526} (\bibinfo {year} {1995})}\BibitemShut
  {NoStop}%
\bibitem [{\citenamefont {Girard}(2005)}]{girard_near_2005}%
  \BibitemOpen
  \bibfield  {author} {\bibinfo {author} {\bibfnamefont {C.}~\bibnamefont
  {Girard}},\ }\href {\doibase 10.1088/0034-4885/68/8/R05} {\bibfield
  {journal} {\bibinfo  {journal} {Reports on Progress in Physics}\ }\textbf
  {\bibinfo {volume} {68}},\ \bibinfo {pages} {1883} (\bibinfo {year}
  {2005})}\BibitemShut {NoStop}%
\bibitem [{\citenamefont {Agarwal}(1975)}]{agarwal_quantum_1975}%
  \BibitemOpen
  \bibfield  {author} {\bibinfo {author} {\bibfnamefont {G.~S.}\ \bibnamefont
  {Agarwal}},\ }\href {\doibase 10.1103/PhysRevA.11.230} {\bibfield  {journal}
  {\bibinfo  {journal} {Physical Review A}\ }\textbf {\bibinfo {volume} {11}},\
  \bibinfo {pages} {230} (\bibinfo {year} {1975})}\BibitemShut {NoStop}%
\bibitem [{\citenamefont {Sersic}\ \emph {et~al.}(2011)\citenamefont {Sersic},
  \citenamefont {Tuambilangana}, \citenamefont {Kampfrath},\ and\ \citenamefont
  {Koenderink}}]{sersic_magnetoelectric_2011}%
  \BibitemOpen
  \bibfield  {author} {\bibinfo {author} {\bibfnamefont {I.}~\bibnamefont
  {Sersic}}, \bibinfo {author} {\bibfnamefont {C.}~\bibnamefont
  {Tuambilangana}}, \bibinfo {author} {\bibfnamefont {T.}~\bibnamefont
  {Kampfrath}}, \ and\ \bibinfo {author} {\bibfnamefont {A.~F.}\ \bibnamefont
  {Koenderink}},\ }\href {\doibase 10.1103/PhysRevB.83.245102} {\bibfield
  {journal} {\bibinfo  {journal} {Physical Review B}\ }\textbf {\bibinfo
  {volume} {83}},\ \bibinfo {pages} {245102} (\bibinfo {year}
  {2011})}\BibitemShut {NoStop}%
\bibitem [{\citenamefont {Kwadrin}\ and\ \citenamefont
  {Koenderink}(2013)}]{kwadrin_probing_2013}%
  \BibitemOpen
  \bibfield  {author} {\bibinfo {author} {\bibfnamefont {A.}~\bibnamefont
  {Kwadrin}}\ and\ \bibinfo {author} {\bibfnamefont {A.~F.}\ \bibnamefont
  {Koenderink}},\ }\href {\doibase 10.1103/PhysRevB.87.125123} {\bibfield
  {journal} {\bibinfo  {journal} {Physical Review B}\ }\textbf {\bibinfo
  {volume} {87}},\ \bibinfo {pages} {125123} (\bibinfo {year}
  {2013})}\BibitemShut {NoStop}%
\bibitem [{\citenamefont {Lunnemann}\ and\ \citenamefont
  {Koenderink}(2016)}]{lunnemann_local_2016}%
  \BibitemOpen
  \bibfield  {author} {\bibinfo {author} {\bibfnamefont {P.}~\bibnamefont
  {Lunnemann}}\ and\ \bibinfo {author} {\bibfnamefont {A.~F.}\ \bibnamefont
  {Koenderink}},\ }\href {\doibase 10.1038/srep20655} {\bibfield  {journal}
  {\bibinfo  {journal} {Scientific Reports}\ }\textbf {\bibinfo {volume} {6}},\
  \bibinfo {pages} {srep20655} (\bibinfo {year} {2016})}\BibitemShut {NoStop}%
\bibitem [{\citenamefont {Pendry}(2000)}]{pendry_negative_2000}%
  \BibitemOpen
  \bibfield  {author} {\bibinfo {author} {\bibfnamefont {J.~B.}\ \bibnamefont
  {Pendry}},\ }\href {\doibase 10.1103/PhysRevLett.85.3966} {\bibfield
  {journal} {\bibinfo  {journal} {Physical Review Letters}\ }\textbf {\bibinfo
  {volume} {85}},\ \bibinfo {pages} {3966} (\bibinfo {year}
  {2000})}\BibitemShut {NoStop}%
\bibitem [{\citenamefont {Soukoulis}\ \emph {et~al.}(2007)\citenamefont
  {Soukoulis}, \citenamefont {Linden},\ and\ \citenamefont
  {Wegener}}]{soukoulis_negative_2007}%
  \BibitemOpen
  \bibfield  {author} {\bibinfo {author} {\bibfnamefont {C.~M.}\ \bibnamefont
  {Soukoulis}}, \bibinfo {author} {\bibfnamefont {S.}~\bibnamefont {Linden}}, \
  and\ \bibinfo {author} {\bibfnamefont {M.}~\bibnamefont {Wegener}},\ }\href
  {\doibase 10.1126/science.1136481} {\bibfield  {journal} {\bibinfo  {journal}
  {Science}\ }\textbf {\bibinfo {volume} {315}},\ \bibinfo {pages} {47}
  (\bibinfo {year} {2007})}\BibitemShut {NoStop}%
\bibitem [{\citenamefont {Arbouet}\ \emph {et~al.}(2014)\citenamefont
  {Arbouet}, \citenamefont {Mlayah}, \citenamefont {Girard},\ and\
  \citenamefont {{Colas des Francs}}}]{arbouet_electron_2014}%
  \BibitemOpen
  \bibfield  {author} {\bibinfo {author} {\bibfnamefont {A.}~\bibnamefont
  {Arbouet}}, \bibinfo {author} {\bibfnamefont {A.}~\bibnamefont {Mlayah}},
  \bibinfo {author} {\bibfnamefont {C.}~\bibnamefont {Girard}}, \ and\ \bibinfo
  {author} {\bibfnamefont {G.}~\bibnamefont {{Colas des Francs}}},\ }\href
  {\doibase 10.1088/1367-2630/16/11/113012} {\bibfield  {journal} {\bibinfo
  {journal} {New Journal of Physics}\ }\textbf {\bibinfo {volume} {16}},\
  \bibinfo {pages} {113012} (\bibinfo {year} {2014})}\BibitemShut {NoStop}%
\bibitem [{\citenamefont {Bakker}\ \emph {et~al.}(2015)\citenamefont {Bakker},
  \citenamefont {Permyakov}, \citenamefont {Yu}, \citenamefont {Markovich},
  \citenamefont {Paniagua-Dom{\'\i}nguez}, \citenamefont {Gonzaga},
  \citenamefont {Samusev}, \citenamefont {Kivshar}, \citenamefont
  {Luk'yanchuk},\ and\ \citenamefont {Kuznetsov}}]{bakker_magnetic_2015}%
  \BibitemOpen
  \bibfield  {author} {\bibinfo {author} {\bibfnamefont {R.~M.}\ \bibnamefont
  {Bakker}}, \bibinfo {author} {\bibfnamefont {D.}~\bibnamefont {Permyakov}},
  \bibinfo {author} {\bibfnamefont {Y.~F.}\ \bibnamefont {Yu}}, \bibinfo
  {author} {\bibfnamefont {D.}~\bibnamefont {Markovich}}, \bibinfo {author}
  {\bibfnamefont {R.}~\bibnamefont {Paniagua-Dom{\'\i}nguez}}, \bibinfo
  {author} {\bibfnamefont {L.}~\bibnamefont {Gonzaga}}, \bibinfo {author}
  {\bibfnamefont {A.}~\bibnamefont {Samusev}}, \bibinfo {author} {\bibfnamefont
  {Y.}~\bibnamefont {Kivshar}}, \bibinfo {author} {\bibfnamefont
  {B.}~\bibnamefont {Luk'yanchuk}}, \ and\ \bibinfo {author} {\bibfnamefont
  {A.~I.}\ \bibnamefont {Kuznetsov}},\ }\href {\doibase
  10.1021/acs.nanolett.5b00128} {\bibfield  {journal} {\bibinfo  {journal}
  {Nano Letters}\ }\textbf {\bibinfo {volume} {15}},\ \bibinfo {pages} {2137}
  (\bibinfo {year} {2015})}\BibitemShut {NoStop}%
\bibitem [{\citenamefont {Decker}\ and\ \citenamefont
  {Staude}(2016)}]{decker_resonant_2016}%
  \BibitemOpen
  \bibfield  {author} {\bibinfo {author} {\bibfnamefont {M.}~\bibnamefont
  {Decker}}\ and\ \bibinfo {author} {\bibfnamefont {I.}~\bibnamefont
  {Staude}},\ }\href {\doibase 10.1088/2040-8978/18/10/103001} {\bibfield
  {journal} {\bibinfo  {journal} {Journal of Optics}\ }\textbf {\bibinfo
  {volume} {18}},\ \bibinfo {pages} {103001} (\bibinfo {year}
  {2016})}\BibitemShut {NoStop}%
\bibitem [{\citenamefont {Kuznetsov}\ \emph {et~al.}(2016)\citenamefont
  {Kuznetsov}, \citenamefont {Miroshnichenko}, \citenamefont {Brongersma},
  \citenamefont {Kivshar},\ and\ \citenamefont
  {Luk'yanchuk}}]{kuznetsov_optically_2016}%
  \BibitemOpen
  \bibfield  {author} {\bibinfo {author} {\bibfnamefont {A.~I.}\ \bibnamefont
  {Kuznetsov}}, \bibinfo {author} {\bibfnamefont {A.~E.}\ \bibnamefont
  {Miroshnichenko}}, \bibinfo {author} {\bibfnamefont {M.~L.}\ \bibnamefont
  {Brongersma}}, \bibinfo {author} {\bibfnamefont {Y.~S.}\ \bibnamefont
  {Kivshar}}, \ and\ \bibinfo {author} {\bibfnamefont {B.}~\bibnamefont
  {Luk'yanchuk}},\ }\href {\doibase 10.1126/science.aag2472} {\bibfield
  {journal} {\bibinfo  {journal} {Science}\ }\textbf {\bibinfo {volume} {354}}
  (\bibinfo {year} {2016}),\ 10.1126/science.aag2472}\BibitemShut {NoStop}%
\bibitem [{\citenamefont {Taminiau}\ \emph {et~al.}(2012)\citenamefont
  {Taminiau}, \citenamefont {Karaveli}, \citenamefont {van Hulst},\ and\
  \citenamefont {Zia}}]{taminiau_quantifying_2012}%
  \BibitemOpen
  \bibfield  {author} {\bibinfo {author} {\bibfnamefont {T.~H.}\ \bibnamefont
  {Taminiau}}, \bibinfo {author} {\bibfnamefont {S.}~\bibnamefont {Karaveli}},
  \bibinfo {author} {\bibfnamefont {N.~F.}\ \bibnamefont {van Hulst}}, \ and\
  \bibinfo {author} {\bibfnamefont {R.}~\bibnamefont {Zia}},\ }\href {\doibase
  10.1038/ncomms1984} {\bibfield  {journal} {\bibinfo  {journal} {Nature
  Communications}\ }\textbf {\bibinfo {volume} {3}},\ \bibinfo {pages}
  {ncomms1984} (\bibinfo {year} {2012})}\BibitemShut {NoStop}%
\bibitem [{\citenamefont {Li}\ \emph {et~al.}(2014)\citenamefont {Li},
  \citenamefont {Jiang}, \citenamefont {Cueff}, \citenamefont {Dodson},
  \citenamefont {Karaveli},\ and\ \citenamefont {Zia}}]{li_quantifying_2014}%
  \BibitemOpen
  \bibfield  {author} {\bibinfo {author} {\bibfnamefont {D.}~\bibnamefont
  {Li}}, \bibinfo {author} {\bibfnamefont {M.}~\bibnamefont {Jiang}}, \bibinfo
  {author} {\bibfnamefont {S.}~\bibnamefont {Cueff}}, \bibinfo {author}
  {\bibfnamefont {C.~M.}\ \bibnamefont {Dodson}}, \bibinfo {author}
  {\bibfnamefont {S.}~\bibnamefont {Karaveli}}, \ and\ \bibinfo {author}
  {\bibfnamefont {R.}~\bibnamefont {Zia}},\ }\href {\doibase
  10.1103/PhysRevB.89.161409} {\bibfield  {journal} {\bibinfo  {journal}
  {Physical Review B}\ }\textbf {\bibinfo {volume} {89}},\ \bibinfo {pages}
  {161409} (\bibinfo {year} {2014})}\BibitemShut {NoStop}%
\bibitem [{\citenamefont {Wiecha}\ \emph {et~al.}(2017)\citenamefont {Wiecha},
  \citenamefont {Arbouet}, \citenamefont {Girard}, \citenamefont {Lecestre},
  \citenamefont {Larrieu},\ and\ \citenamefont
  {Paillard}}]{wiecha_evolutionary_2017}%
  \BibitemOpen
  \bibfield  {author} {\bibinfo {author} {\bibfnamefont {P.~R.}\ \bibnamefont
  {Wiecha}}, \bibinfo {author} {\bibfnamefont {A.}~\bibnamefont {Arbouet}},
  \bibinfo {author} {\bibfnamefont {C.}~\bibnamefont {Girard}}, \bibinfo
  {author} {\bibfnamefont {A.}~\bibnamefont {Lecestre}}, \bibinfo {author}
  {\bibfnamefont {G.}~\bibnamefont {Larrieu}}, \ and\ \bibinfo {author}
  {\bibfnamefont {V.}~\bibnamefont {Paillard}},\ }\href {\doibase
  10.1038/nnano.2016.224} {\bibfield  {journal} {\bibinfo  {journal} {Nature
  Nanotechnology}\ }\textbf {\bibinfo {volume} {12}},\ \bibinfo {pages} {163}
  (\bibinfo {year} {2017})}\BibitemShut {NoStop}%
\bibitem [{\citenamefont {Johnson}\ and\ \citenamefont
  {Christy}(1972)}]{johnson_optical_1972}%
  \BibitemOpen
  \bibfield  {author} {\bibinfo {author} {\bibfnamefont {P.~B.}\ \bibnamefont
  {Johnson}}\ and\ \bibinfo {author} {\bibfnamefont {R.~W.}\ \bibnamefont
  {Christy}},\ }\href {\doibase 10.1103/PhysRevB.6.4370} {\bibfield  {journal}
  {\bibinfo  {journal} {Physical Review B}\ }\textbf {\bibinfo {volume} {6}},\
  \bibinfo {pages} {4370} (\bibinfo {year} {1972})}\BibitemShut {NoStop}%
\bibitem [{\citenamefont {Girard}\ \emph {et~al.}(2015)\citenamefont {Girard},
  \citenamefont {Cuche}, \citenamefont {Dujardin}, \citenamefont {Arbouet},\
  and\ \citenamefont {Mlayah}}]{girard_molecular_2015}%
  \BibitemOpen
  \bibfield  {author} {\bibinfo {author} {\bibfnamefont {C.}~\bibnamefont
  {Girard}}, \bibinfo {author} {\bibfnamefont {A.}~\bibnamefont {Cuche}},
  \bibinfo {author} {\bibfnamefont {E.}~\bibnamefont {Dujardin}}, \bibinfo
  {author} {\bibfnamefont {A.}~\bibnamefont {Arbouet}}, \ and\ \bibinfo
  {author} {\bibfnamefont {A.}~\bibnamefont {Mlayah}},\ }\href {\doibase
  10.1364/OL.40.002116} {\bibfield  {journal} {\bibinfo  {journal} {Optics
  Letters}\ }\textbf {\bibinfo {volume} {40}},\ \bibinfo {pages} {2116}
  (\bibinfo {year} {2015})}\BibitemShut {NoStop}%
\end{thebibliography}
